\documentclass[aps, prb, reprint,
superscriptaddress, amsmath, tightenlines,
]{revtex4-1}

\usepackage{graphicx}
\usepackage{amsfonts}
\usepackage{amssymb}
\usepackage{amsmath}
\usepackage{lineno}
\usepackage{xcolor}
\usepackage{setspace}
\usepackage[normalem]{ulem}

\newcommand{\ket}[1]{\vert#1\rangle}
\newcommand{\bra}[1]{\langle #1\vert}

\usepackage{xcolor}

\begin{document}
    
\title{
Multiphoton non-local quantum interference controlled by an undetected photon\\
}


\author{Kaiyi Qian}
\affiliation{National Laboratory of Solid-state Microstructures, School of Physics, Collaborative Innovation Center of Advanced Microstructures, Nanjing University, Nanjing 210093, China}

\author{Kai Wang}
\email{kai.wang@nju.edu.cn}
\affiliation{National Laboratory of Solid-state Microstructures, School of Physics, Collaborative Innovation Center of Advanced Microstructures, Nanjing University, Nanjing 210093, China}

\author{Leizhen Chen}
\affiliation{National Laboratory of Solid-state Microstructures, School of Physics, Collaborative Innovation Center of Advanced Microstructures, Nanjing University, Nanjing 210093, China}

\author{Zhaohua Hou}
\affiliation{National Laboratory of Solid-state Microstructures, School of Physics, Collaborative Innovation Center of Advanced Microstructures, Nanjing University, Nanjing 210093, China}

\author{Mario Krenn}
\email{mario.krenn@mpl.mpg.de}
\affiliation{Max Planck Institute for the Science of Light (MPL), Erlangen, Germany}
    
\author{Shining Zhu}
\affiliation{National Laboratory of Solid-state Microstructures, School of Physics, Collaborative Innovation Center of Advanced Microstructures, Nanjing University, Nanjing 210093, China}

\author{Xiao-Song Ma}
\email{Xiaosong.Ma@nju.edu.cn}
\affiliation{National Laboratory of Solid-state Microstructures, School of Physics, Collaborative Innovation Center of Advanced Microstructures, Nanjing University, Nanjing 210093, China}
\affiliation{Synergetic Innovation Center of Quantum Information and Quantum Physics, University of Science and Technology of China, Hefei, Anhui 230026, China}
\affiliation{Hefei National Laboratory, Hefei 230088, China}

\date{\today}
    
\begin{abstract}
The interference of quanta lies at the heart of quantum physics. The multipartite generalization of single-quanta interference creates entanglement, the coherent superposition of states shared by several quanta. Entanglement allows non-local correlations between many quanta and hence is a key resource for quantum information technology. Entanglement is typically considered to be essential for creating non-local quantum interference. Here, we show that this is not the case and demonstrate multiphoton non-local quantum interference that does not require entanglement of any intrinsic properties of the photons. We harness the superposition of the physical origin of a four-photon product state, which leads to constructive and destructive interference with the photons' mere existence. With the intrinsic indistinguishability in the generation process of photons, we realize four-photon frustrated quantum interference. This allows us to observe the following noteworthy difference to quantum entanglement: We control the non-local multipartite quantum interference with a photon that we never detect, which does not require quantum entanglement. These non-local properties pave the way for the studies of foundations of quantum physics and potential applications in quantum technologies.
\end{abstract}
\maketitle
\onecolumngrid
\setstretch{1.7}
\section*{Introduction}
Quantum interference occurs only when no information to distinguish between the superposed states is knowable\cite{RevModPhys.71.S274}. Well-known examples of quantum interference with photons include double-slit interference of a single photon\cite{grangier1986experimental} and Hong-Ou-Mandel interference of two photons\cite{PhysRevLett.59.2044}. A separate type of quantum interference is the interference via induced coherence, first realized by Zou, Wang, and Mandel in 1991\cite{PhysRevLett.67.318,PhysRevA.44.4614} in a Mach-Zehnder interferometer-like configuration. The interference of the signal photon depends on the path identity of its twin photon, which is not even on the coherent paths of the signal photon. This mind-boggling experiment ``brings out that the quantum state reflects not what we know about the system, but rather what is knowable in principle" \cite{RevModPhys.71.S274}. In 1994, Herzog et al. demonstrated frustrated two-photon creation via induced coherence in a Michelson interferometer-like configuration, in which they can either enhance or suppress the generation of photon pairs in spontaneous parametric down-conversion (SPDC) process by tuning the phases of various interferometers\cite{PhysRevLett.72.629}. Throughout this manuscript, we call this type of interference frustrated interference (FI). 

Nonlocality is the characteristic feature of quantum correlation, such as entanglement. For instance, two space-like separated observers --- Alice ($\mathcal{A}$) and Bob ($\mathcal{B}$), share a pair of polarization-entangled photons and measure on specific polarization bases by adjusting the transmission angles ($\alpha/\beta$) of their polarizers (Fig. 1a). When they compare their results, they will find that the joint probability depends on the polarizers' angles: $P_{\mathcal{A}\mathcal{B}}(\alpha+\beta)=\sin^2(\alpha+\beta)$, as shown in Fig. 1a. The probability here is normalized with the maximum counting rate for all the possible measurement settings of $\alpha$ and $\beta$. This second-order interference of the entangled state can not be explained by local hidden variable theory and is considered non-local. In this setting, any mutual influence between the two observations is excluded under strict Einstein locality conditions.\cite{PhysRevLett.81.5039} This phenomenon, predicted by quantum physics, cannot be accounted for by any local theory and represents one of the most profound foundational insights in physics\cite{RevModPhys.86.419}.

In almost all scenarios in which non-local interference is observed, entanglement --- or more generally some form of quantum correlation --- is the basic ingredient. In this work, we show that this is not necessarily the case, and demonstrate multiphoton non-local quantum interference, which does not need entanglement. Note that nonlocality without entanglement has been discussed in the context of quantum state discrimination\cite{PhysRevA.59.1070}, which is not relevant to our work.

In this work we experimentally observe the multiphoton frustrated quantum interference (MFI) --- a concept only theoretically proposed recently\cite{gu2019quantum}. Then we go beyond and demonstrate a surprising physical property: We observe non-local quantum interference that does not require quantum entanglement. Specifically, we tune the phase of a photon that we never detect, and observe interference with the rest photons.

\begin{figure*}
    \includegraphics[width=18cm]{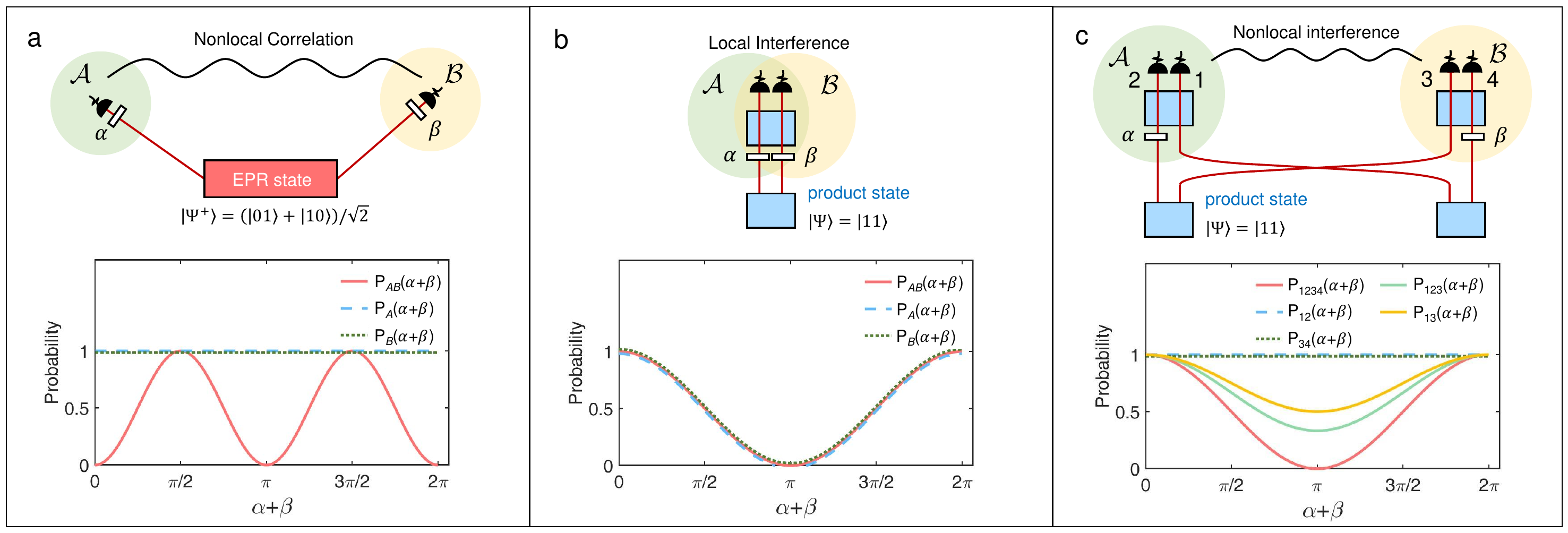}
    \caption{\textbf{Non-local and local quantum interference.}  \textbf{a}, Non-local quantum interference of entangled states. The two photons from one Einstein--Podolsky--Rosen(EPR) state source (red rectangle) have correlations that persist even when two observers measure their respective photons at a distance from each other. The coincidence of Alice ($\mathcal{A}$) and Bob ($\mathcal{B}$) depends on the angles of the polarizer transmission axis, $\alpha$ and $\beta$: $P_{\mathcal{AB}}(\alpha+\beta)=\sin^2(\alpha+\beta)$ (red solid curve in the lower panel of Fig. 1a). All the probabilities in these figures are normalized with the maximum counting rate for all the possible measurement settings. The single counts of $\mathcal{A}$ and $\mathcal{B}$ (blue dash/ green dot curve in Fig. 1a) show no interference when varying $\alpha$ and $\beta$: $P_{\mathcal{A}}(\alpha+\beta)=P_{\mathcal{B}}(\alpha+\beta)=1$. \textbf{b}, Two-photon frustrated interference. One pair of photons is generated from one of the two product-state sources (blue rectangles). These two possible photon-generation processes interfere when they are indistinguishable. The probabilities of detecting two photons (red solid curve in the lower panel of Fig. 1b) and single photons (blue dash/green dot curve in Fig. 1b) depend on the phases in the system: $P_{\mathcal{A}\mathcal{B}}(\alpha+\beta)=P_{\mathcal{A}}(\alpha+\beta)=P_{\mathcal{B}}(\alpha+\beta)=\frac{1}{2}+\frac{1}{2}\cos(\alpha+\beta)$. This type of interference is local, as the detection event of $\mathcal{A}$ will always be in the future light cone of setting event of $\mathcal{B}$($\beta$). \textbf{c}, Four-photon frustrated interference shows a non-local interference, which originates from the indistinguishability of the sources and does not require quantum entanglement. The four-photon coincidence of $\mathcal{A}$ and $\mathcal{B}$ depends on both phases $\alpha$ and $\beta$: $P_{1234}(\alpha+\beta)=\frac{1}{2}+\frac{1}{2}\cos(\alpha+\beta)$ (red solid curve in the lower panel of Fig. 1c). The local measurement of $\mathcal{A}$/$\mathcal{B}$ (two-photon coincidence) shows no interference when varying $\alpha$ and $\beta$: $P_{12}(\alpha+\beta)=P_{34}(\alpha+\beta)=1$ (blue dash/green dot curve in the lower panel of Fig. 1a). The three-fold coincidence count of $\mathcal{A}$ and photon 3 varies with the phase $\beta$ of the undetected photon 4: $P_{123}(\alpha+\beta)=\frac{2}{3}+\frac{1}{3}\cos(\alpha+\beta)$ (light green curve in the lower panel of Fig. 1c). The two-fold coincidence shows the same interference with reduced visibility: $P_{13}(\alpha+\beta)=\frac{3}{4}+\frac{1}{4}\cos(\alpha+\beta)$ (orange curve in the lower panel of Fig. 1c). The settings of $\beta$($\alpha$) can be space-like separated from the detection of $\mathcal{A}$($\mathcal{B}$), as in the case of entangled state (Fig. 1a).}
\end{figure*}

\section*{Results}
\subsection*{The property of two-photon frustrated interference}
To understand the MFI, we first review the two-photon FI, of which the conceptual scheme is given in Fig. 1b. Two down-conversion crystals are coherently pumped and probabilistically generate one photon pair. When we cannot distinguish which crystal the two photons come from, the coincidence of $\mathcal{A}$ and $\mathcal{B}$ oscillates as a function of phase $\beta$: $P_{\mathcal{A}\mathcal{B}}(\alpha+\beta)=\frac{1}{2}+\frac{1}{2}\cos(\alpha+\beta)$ (Fig. 1b). Moreover, FI even does not require the coincidence measurement as in the entanglement scenario. The single counts of A show the interference, depending on a phase $\beta$ with no direct interaction: $P_{\mathcal{A}}(\alpha+\beta)=\frac{1}{2}+\frac{1}{2}\cos(\alpha+\beta)$ (Fig. 1b). This phenomenon is beyond the quantum entanglement, as a subsystem of a maximally entangled state is in a mixed state and shows no interference (Fig. 1a). Profit from this property, there has been a resurgence of interest in applying FI to quantum-enhanced techniques recently, such as quantum imaging\cite{lemos2014quantum}, spectroscopy\cite{kalashnikov2016infrared,paterova2017nonlinear,paterova2018measurement}, optical coherence tomography\cite{paterova2018tunable}, state generation\cite{PhysRevLett.118.080401,su2019versatile}, microscopy\cite{paterova2020hyperspectral}, bio-imaging\cite{kviatkovsky2020microscopy} and quantum holography\cite{topfer2022quantum}. This resurgence is fuelled by the application of non-degenerate photon pairs in FI, where one can probe objects of interest with the longer-wavelength photon, and measure the result with a shorter-wavelength photon that can easily be detected. For details, see the recent review on this topic\cite{hochrainer2021quantum}. Note in the strong squeezing limit, one can use the so-called SU(1,1) interferometer for improving phase sensitivity\cite{PhysRevA.33.4033,chekhova2016nonlinear,ou2020quantum}.

However, this property of two-photon FI shows only the local interference. As shown in Fig. 1b, the two-photon case can not be non-local even in principle. The phase tuning event of the signals and idlers ($\alpha$ and $\beta$) are always in the backward light cones of the detection events. Under strict Einstein's locality condition, a non-local configuration requires that the measurement result of Alice will not be influenced by the measurement setting $\beta$ of Bob within the time that the light travels. As shown in Fig. 1b, the two-photon frustrated interference case\cite{PhysRevLett.67.318,PhysRevA.44.4614,PhysRevLett.72.629}, including quantum imaging with undetected light\cite{lemos2014quantum}, will always be local under enforced Einstein locality conditions\cite{herzog1994herzog}. 

\subsection*{The property of four-photon frustrated interference}
Here, we extend FI to a four-photon case to realize a non-local multiphoton interference. By non-local interference we mean an interferometer where the phase setting and port of the interferometer can be spatially separated under strict Einstein locality conditions (see Supplementary Note 7 for a space-time diagram). We employ four photon-pair sources in a configuration in which only two pairs of product states are generated from them (Fig. 1c). Alice and Bob control their phase shifters ($\alpha/\beta$) locally and measure the four-fold coincidence counts, in which case they receive a product state. Since the settings of $\mathcal{A}$ ($\mathcal{B}$) can be space-like separated from detection events of $\mathcal{B}$($\mathcal{A}$), they obtain the non-local phase-dependent coincidence counts, that is, the four-fold coincidence counts oscillate as a function of $\alpha/\beta$ (Fig. 1c). Therefore, we call the four-photon FI non-local quantum interference, as the photon-count dependence between $\mathcal{A}$ and $\mathcal{B}$ still remains, even if they are space-like separated.

This measurement with the product state is very similar to the non-local interference with entangled states. However, here no quantum entanglement between any properties of the photons exists, but one can observe interference with the mere existence of a multi-photon state. It arrives from a coherent superposition of the origin of the multi-photon state. Moreover, when Bob varies the phase $\beta$ and measures the three-fold coincidence between the two detectors in $\mathcal{A}$ and detector 3 in $\mathcal{B}$ (Fig. 1c), they will observe the interference of the three photons as a function of $\beta$ (Fig. 1c). We stress that the phase $\beta$, which we can tune, has no direct interaction with all the other three detected photons. This is the unique feature of MFI and in contrast to the entanglement case, where $P_{\mathcal{A}}(\alpha+\beta)$ does not depend on $\beta$ (Fig. 1a). Although the visibility of $P_{123}(\alpha+\beta)$ is not 1 due to the particular construction of the setup, more complex source configurations and detection schemes may further increase the visibility. In this case, we can probe the three-photon coincidence count by tuning the phase of the fourth photon, which is undetected. We note that one cannot achieve space-like separation between the detection on $\mathcal{A}$, photon 3, and the setting $\beta$. Otherwise, superluminal control would occur.

From a fundamental perspective, by extending the two-photon FI to multiphoton FI, one could separate the down-conversion crystals in space and demonstrate non-local control of multiphoton interference that does not need entanglement. From an application perspective, one could devise more complex quantum-information tasks, such as quantum computation\cite{gu2019quantum} and generations of complex multi-photon quantum states\cite{PhysRevLett.118.080401,lu2020three,PhysRevX.11.031044}.

\subsection*{Scheme of four-photon frustrated interference}
In this work, four photons are generated in two indistinguishable generation processes and measured with four detectors, enabling the suppression and enhancement of four-photon generation via FI\cite{PhysRevLett.119.240403,gu2019quantum}. In two-photon FI, there is one pair of photons generated from two two-photon sources\cite{PhysRevLett.72.629}. In the four-photon FI demonstrated here, we use four two-photon sources for generating two pairs of correlated photons, as shown in Fig. 2a. Four two-photon sources placed in sequence are pumped coherently by two laser beams. The down-converted photons from different groups (crystals I and II, and crystals III and IV) are aligned according to the geometry shown in Fig. 2a to ensure the path indistinguishability. Photons on the same path have identical properties (such as polarization, frequency, and arrival time at the detectors). We emphasize that there is no entanglement of any external or internal degrees of freedoms of photons involved, neither those created by the source (as we use single-mode fibers) nor those generated through the concept of entanglement by path identity (as we do not shift modes between pair creations, which is the key idea of entanglement by path identity\cite{PhysRevLett.118.080401}).

\begin{figure*}
    \includegraphics[width=16cm]{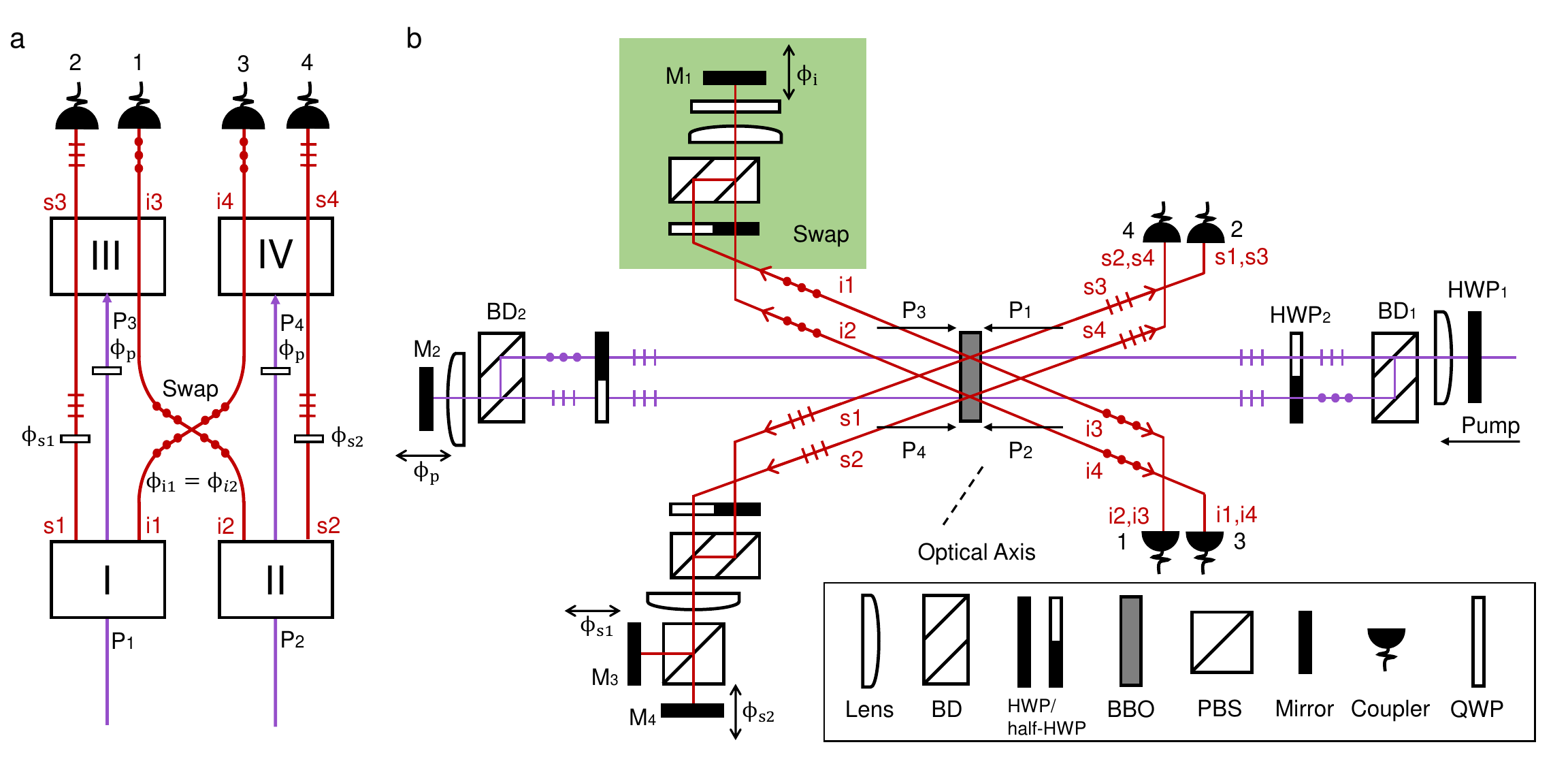}
    \caption{\textbf{Four-photon frustrated interference.} \textbf{a}, Scheme of frustrated four-photon interference. Four-fold coincidence events occur when crystals I and II, or crystals III and IV generate two pairs of photons simultaneously. $\phi_{sX}$ and $\phi_{iX}$ represent the phase of the signal and idler photon from crystal X, respectively, and $\phi_p$ is the phase of the pumps. The interference pattern emerges when we cannot distinguish which group the four photons come from. The quantum state is given by $\ket{\psi}=[e^{i(\phi_{i1}+\phi_{s1}+\phi_{i2}+\phi_{s2})}+e^{i2\phi_p}]\ket{1111}$, which is a product state and has no entanglement. \textbf{b}, Experimental setup. The pump incidents from the right side and splits on a beam displacer (BD1) to generate the four photons via SPDC in a `back-reflect' configuration, where the phase $\phi_p$ is controlled by M2. The idlers of sources I and II (i1, i2) exchange their path by polarization in the Swap module. Therefore, i1 and i2 experience the same phase $\phi_i$. The phases of signals (s1, s2) are controlled independently by M3 and M4, respectively. I1, i2, s1, and s2 are aligned with i3, i4, s3, and s4, respectively, ensuring the path identity.  All four photons are finally collected by couplers 1--4 and detected with single-photon detectors. See main text for details. }
\end{figure*}

Considering the low probability $p$ for generating photon pairs for the SPDC process, the output state (without normalization) from modes 1 to 4 can be written as:

\vspace{-0.3cm}
\begin{align}
\nonumber \ket{\psi}=&\ket{{\rm vac}}+p[e^{i(\phi_{s1}+\phi_{i1})}\ket{0110}+e^{i(\phi_{s2}+\phi_{i2})}\ket{1001}+e^{i\phi_{p}}\ket{1100}+e^{i\phi_{p}}\ket{0011}]\\
\nonumber &+p^2[e^{i(\phi_{i1}+\phi_{s1}+\phi_{i2}+\phi_{s2})}\ket{1111}+e^{i2\phi_p}\ket{1111}\\
&+\sqrt{2}e^{i(\phi_{i1}+\phi_{s1}+\phi_p)}\ket{1210}+\sqrt{2}e^{i(\phi_{i1}+\phi_{s1}+\phi_p)}\ket{0121}+\sqrt{2}e^{i(\phi_p+\phi_{i2}+\phi_{s2})}\ket{2101}+\sqrt{2}e^{i(\phi_p+\phi_{i2}+\phi_{s2})}\ket{1012}]
\end{align}
to second-order approximation, where the numbers in the kets represent photon numbers in modes 1 to 4. 

When we count only the event detecting the four photons simultaneously, we post-select the state $\ket{1111}$, which is a product state, as shown in the second line of Eq. (1). The probability of observing four-fold coincidence counts $(P_{1,2,3,4})$ varies with the phases in the system:
\begin{equation}
P_{1,2,3,4}=p^4[2+2\cos{(\phi_{i1}+\phi_{s1}+\phi_{i2}+\phi_{s2}-2\phi_p)}].
\end{equation}
See Supplementary Note 1 for detailed derivation.

A more interesting phenomenon, multiphoton interference controlled by an undetected photon, is observed in our experiment. When we vary the phase of photon s1, $\phi_{s1}$, the probability of observing the other three photons detected by detectors 1, 3, and 4 is:
\begin{equation}
P_{1,3,4}=p^4[4+2\cos{(\phi_{i1}+\phi_{s1}+\phi_{i2}+\phi_{s2}-2\phi_p)}].
\end{equation}
The ideal visibility is 50\% due to the multiphoton noise from $\ket{1012}$. In this case, we can detect and observe three-photon interference by tuning the phase of the fourth photon, which is undetected. This \sout{novel} finding shows the effect that one has non-local quantum interference that does not require entangled states. For entangled states, such as a Greenberger--Horne--Zeilinger(GHZ) state\cite{greenberger1989going,greenberger1990bell}, one would lose correlation when losing one particle.

\subsection*{Experimental setup}
The scheme of our experimental setup is shown in Fig. 2b. The dimension of the optical setup is roughly $0.8 \times 1.0$~$\rm m^2$. The pump is a 404-nm femtosecond pulsed laser with vertical polarization.  A half-wave plate (HWP1) rotates the polarization of the pump laser to $45^{\circ}$. One polarization beam displacer (BD1) separates the pump laser into two parallel paths with equal power of about 0.29~W, denoted as P1 (H) and P2 (V), to pump a single beta-barium borate (BBO) crystal separately. The spacing between the two paths is about 4~mm. Both P1 and P2 are horizontal polarization after the semicircle HWP2 (half-HWP). The optical axis of BBO is in the horizontal plane and is aligned to be $40.9^{\circ}$ with respect to the two pumps to form the beamlike SPDC configuration\cite{Niu:08,PhysRevLett.117.210502}. 

P1 and P2 generate two pairs of photons denoted as s1, i1, and s2, i2. The photon pairs from the beamlike source are in the polarization product state $\rm \ket{HV}_{si}$, and the emission angles of signal and idler with respect to the pump are approximately $3^{\circ}$. The polarization of the down-converted photons is shown in Fig. 2b. The triple dots represent vertical polarization and the triple lines represent horizontal polarization. As s1/i1 is parallel with s2/i2, after the semicircle HWP, both s1 (V), s2 (H), and i1 (V), i2 (H) are combined on the BDs and are focused with lenses to improve the coupling efficiency at the couplers. Photons i1 and i2 pass through a quarter wave plate (QWP) with the angle fixed at $45^{\circ}$. Then they are reflected on mirror M1. Therefore, the two photons swap their path on the way back, which corresponds to the crossing between photons i1 and i2 in Fig. 2a. On the signal photons side, we separate s1 and s2 on a polarization beam splitter (PBS) to control their phase $\phi_{s1}$ and $\phi_{s2}$ independently.
 
After the BBO crystal, P1 and P2 are combined on BD2 and reflected by mirror M2, forming a symmetrical interferometer. The reflected pumps, denoted as P3 and P4, are used to generate photon pairs s3 and i3, and s4 and i4. By adjusting M1, M3, and M4, the paths of s1, i1, s2, and i2 overlap with s3, i4, s4, and i3, respectively, as shown in Fig. 2a, which erases the path distinguishability. Though the polarization states of signals and idlers are different, they are the same for the photons on the same path due to the symmetry of our interferometer, which is necessary for realizing the four-photon interference. To observe the four-photon interference successfully, we also need to erase the temporal distinguishability. We fix M2 and scan the delays of M1, M3, and M4 until the interference pattern emerges, ensuring that the reflected photons and the reflected pump laser pulses arrive at the crystal simultaneously. We note that there is a time difference between the reflected pumps P3 and P4 due to the geometrical dimension of the BDs, and so are the signals and idlers on the same side. We can still realize the four-photon interference. We only need to ensure photons on the same path arrive simultaneously, not all the photons on different paths\cite{PhysRevLett.77.1917}. This is especially important for future space-like separated experiments of this effect. For the detailed results of path identity and analysis of timing, see Supplementary Note 4.

All four photons s1 (s3), s2 (s4), i1 (i4), and i2 (i3) are finally collected by single-mode fiber couplers. We analyze the coincidence counts while varying the phase $\phi_{s1}$ of s1. The result of four-fold coincidence counts is shown in Fig. 3a. The period of the interference pattern is 403.5~nm, in agreement with the 808-nm central wavelength of photon s1, considering that it goes back and forth. The visibility of interference is about 75.47\%. The misalignment of photons on the same path reduces the identity in spatial mode and thereby the four-photon interference visibility. Based on the values obtained from independent experimental measurements, the estimated maximum achievable value for the visibility is about 81.95\% (see Supplementary Note 3), which is higher than we obtained (75.47\%). This discrepancy may come from higher-order emission from SPDC, which further reduces the four-fold interference visibility.
\begin{figure*}
    \includegraphics[width=16cm]{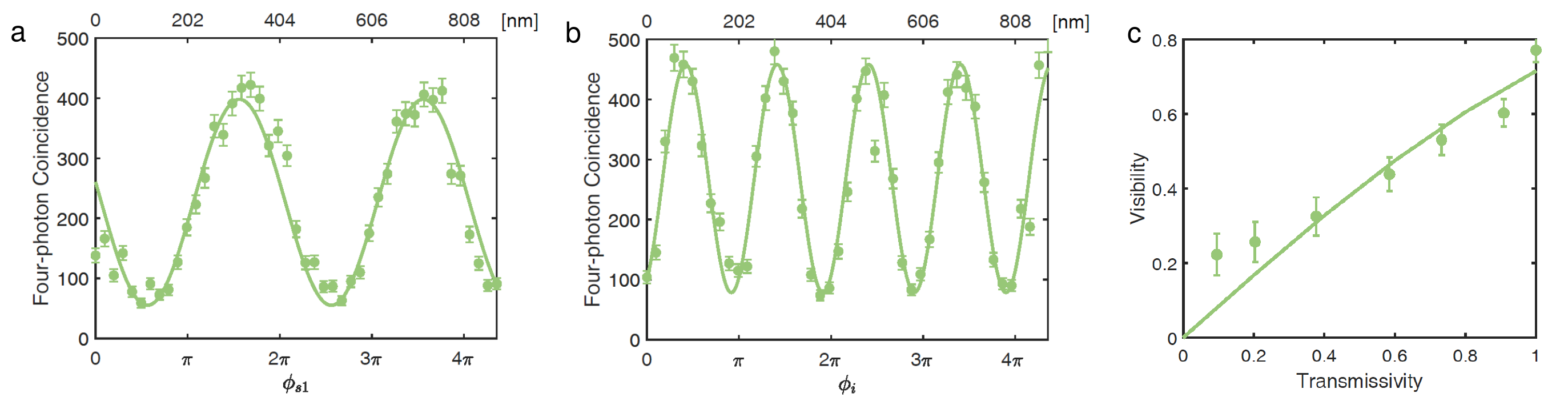}
    \caption{\textbf{Results of four-fold coincidence counts for multiphoton frustrated interference.} \textbf{a}, The horizontal axis represents the position of M3 ($\phi_{s1}$). The interference pattern of frustrated four-photon interference has visibility of $75.47\%\pm2.99\%$ and a period of 403.5 nm. The green line is the fitting curve. \textbf{b}, The horizontal axis represents the position of M1 ($\phi_i$). The interference pattern has visibility of $74.26\%\pm2.79\%$ and a period of 200.9 nm. The errors of visibilities are derived from Poisson statistics. The integration time for each point in \textbf{a} and \textbf{b} is 30 s. \textbf{c}, The relationship between the visibility of four-photon coincidence and the transmissivity of photon s2. The green line is a fit of the data points according to function $V=\frac{2\alpha T}{1+\alpha^2T^2}$, with $\alpha=0.42$. All the data presented in this manuscript are the raw data with no noise subtraction.
     }
\end{figure*}

The spatial misalignment causes experimental visibility different from identity. It can be modeled by including the transmissivity ($T$) in the path of photon s2 [see ref 4]. Therefore, we reduce the coupling efficiency of photon s2 (hence lower $T$) and measure the visibility of four-fold coincidence to verify this effect. We note that the visibility is not an exact linear correlation in $T$ for four-fold coincidence: $V=\frac{2\alpha T}{1+\alpha^2T^2}$ (see Supplementary Note 6), where $\alpha$ is the parameter used to characterize the path identity. The experimental result is shown in Fig. 3c. As the transmissivity of photon s2 decreases, the visibility of interference goes down to almost zero. That is because we know that photons on mode-4 come from crystal IV when s2 is blocked, which destroys the interference. The parameter $\alpha$ of the fitting curve is 0.42.

We also scan phase $\phi_i$ and record the four-fold coincidence counts. The result is shown in Fig. 3b. Because both signals i1 and i2 experience phase $\phi_i$ as in Eq. (2), the interference period is approximately 200.9 nm, which is half of the period shown in Fig. 3a. The visibility of the interference is 74.26\%. It is consistent with the visibility of Fig. 3a.

\begin{figure*}
    \includegraphics[width=16cm]{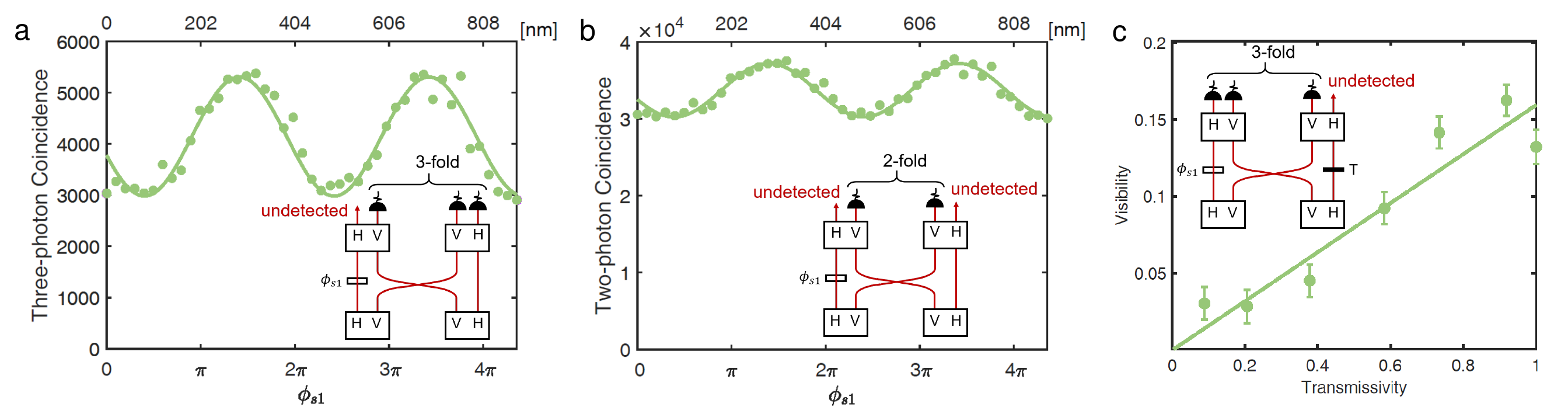}
    \caption{\textbf{Multi-photon non-local frustrated interference with undetected photons.}  \textbf{a}, Result of three-fold coincidence counts on detectors 1, 3 and 4. The horizontal axis represents the position of M3 ($\phi_{s1}$). The interference pattern has visibility of $29.84\%\pm1.05\%$ and a period of 407.1~nm, almost the same as in Fig. 3a. \textbf{b}, Two-fold coincidence counts on detectors 1 and 3. The interference pattern has visibility of $11.42\%\pm0.38\%$ and a period of 407.2nm. The errors of visibilities are derived from Poisson statistics. The integration time for each point in \textbf{a} and \textbf{b} is 30 s. \textbf{c}, The relationship between the visibility of three-photon coincidence (detector 1, 2, 3) and the transmissivity of photon s2. The green line is a linear fit to the data points. All the data presented in this manuscript are the raw data with no noise subtraction.
     }
\end{figure*}

\subsection*{Multi-photon frustrated interference controlled by an undetected photon}
To demonstrate the multiphoton interference controlled by an undetected photon in the frustrated interference, we change the phase of s1 and measure the three-fold coincidence events on detectors 1, 3, and 4, where photon s1 is undetected (see the inset of Fig. 4a). The result is shown in Fig. 4a. As the phase $\phi_{s1}$ varies, the coincidence counts of the other three photons change correspondingly. Therefore, we observe multipartite frustrated interference, where multiple correlated photons are influenced by a phase that has no direct relevance. The visibility of the interference is 29.84\%, which is lower than the theoretical value of 50\%. The limited visibility is because of the limited path indistinguishability for idler photons. We furthermore show that even two-fold coincidence, on detectors 1 and 3 (see the inset of Fig. 4b), can be controlled with the phase of undetected photon s1, $\phi_{s1}$: 
\begin{equation}
P_{1,3}=p^4[6+2\cos{(\phi_{i1}+\phi_{s1}+\phi_{i2}+\phi_{s2}-2\phi_p)}].
\end{equation}
The result of the coincidence measurement on detectors 1 and 3 is shown in Fig. 4b. As the coincidence only occurs when more than two crystals generate photons, twin photons from the same crystal do not cover up the interference. The interference data in Figs. 3a and 4a, b are recorded simultaneously and show nearly identical interference phase-dependence. 

Finally, to show that our experiment is a genuine quantum mechanical effect and a consequence of induced coherence, we vary the phase $\phi_{s1}$ and measure the interference visibility of three-photon coincidence on detectors 1, 2, and 3 while reducing the transmissivity of photon s2, as shown in Fig. 4c. The nearly linear relation indicates that the four-photon FI is an induced coherence rather than induced stimulation\cite{wiseman2000induced,PhysRevA.100.053839}, which is beyond the classical optics\cite{PhysRevLett.67.318}.

\section*{Discussion}
In this work, by harnessing the indistinguishability between the generation process of photons, we have shown four-photon non-local quantum interference with product states. This effect occurs not because of a superposition of the photons' external properties, such as path, polarization, and so on. Instead, it happens because of a fundamental unknowability where the photons have been generated. This underlying principle allows us to show how we can manipulate the interference of three photons by introducing a phase in the fourth photon that we never detect. Other types of interference: single-photon interference, and two-photon interference (entanglement and Hong-Ou-Mandel), all require the detection of all involved photons to observe the interference effect. If one traces out one photon, the outcomes show no interference. For instance, if one uses two identical single photons generated from two independent sources and performs the HOM interference experiment, one will not see any interference effect if one only measures the single-photon counts. Interference only appears when the coincidence measurements of all interfering photons are performed. In this work, the situation is fundamentally different. We tune the phase of a photon that we never detect, and observe interference of the rest photons. Multiphoton interference appears when the coincidence measurements of only partial interfering photons are performed.

Also, this is a different interference phenomenon to some of the maximally entangled states, such as GHZ states. If one photon of a GHZ state remains undetected, no quantum interference will be observed in the rest photons (see Supplementary Note 9 for details). Our experiment thereby demonstrates in a direct way how the lack of knowledge about a quantum system can lead to multiphoton non-local quantum interference, a feature that cannot solely be described by entanglement.

Novel properties of this quantum system can be observed with improvements in our experimental setup. As we purposefully chose to build our setup with bulk optics, we can separate the distance between the crystals and measure the non-local interference influenced by the phase of the undetected photon, which is important for exploring fundamental questions of quantum physics and may be useful in quantum communication. This is possible because one can build an experimental setup that shows the nonlocal interference under strict Einstein locality conditions, by randomly setting the phases $\alpha$ or $\beta$ (in Fig. 1c) after photon pairs in the lower layer (more details in Supplementary Information).

Additionally, variations of our multi-photon experiments with induced coherence can be used to explore highly diverse quantum systems. An example is a resource state for photonic quantum computers\cite{rudolph2017optimistic}, in which one exploits the exact multi-photon frustrated generation that we experimentally showed here.

Finally, the demonstration of non-local interference with undetected photons relates our experiment to a very vibrant field of quantum imaging with undetected photons\cite{lemos2014quantum} and its variations\cite{hochrainer2021quantum}. Here, one striking property is the generation of multiple wavelengths of the different photons. This is interesting in the absence of suitable detectors for the wavelength of the undetected photon. Our experiment brings this application-driven research finally into the multi-photon regime\cite{lemos2014quantum,kalashnikov2016infrared,paterova2017nonlinear,paterova2018measurement,kviatkovsky2020microscopy,paterova2020hyperspectral,paterova2018tunable}. In this work, we have shown the multi-photon frustrated interference, which can potentially be observed even when the settings and detections of Alice and Bob are space-like separated. Note that this unique property of multiphoton frustrated interference was not discussed in the original proposal of multi-photon frustrated interference\cite{gu2019quantum} nor was it shown in a recent related work\cite{Feng:23}. 


\section*{Data availability}
The data that support the plots within this paper and other findings of this study are available at https://github.com/NJU-Malab/Frustrated-Interference\href{https://github.com/NJU-Malab/Frustrated-Interference}.

%
%


\section*{References}

    %
    %

\begin{thebibliography}{10}
\expandafter\ifx\csname url\endcsname\relax
  \def\url#1{\texttt{#1}}\fi
\expandafter\ifx\csname urlprefix\endcsname\relax\def\urlprefix{URL }\fi
\providecommand{\bibinfo}[2]{#2}
\providecommand{\eprint}[2][]{\url{#2}}

\bibitem{RevModPhys.71.S274}
\bibinfo{author}{Mandel, L.}
\newblock \bibinfo{title}{Quantum effects in one-photon and two-photon
  interference}.
\newblock \emph{\bibinfo{journal}{Rev. Mod. Phys.}}
  \textbf{\bibinfo{volume}{71}}, \bibinfo{pages}{S274--S282}
  (\bibinfo{year}{1999}).

\bibitem{grangier1986experimental}
\bibinfo{author}{Grangier, P.}, \bibinfo{author}{Roger, G.} \&
  \bibinfo{author}{Aspect, A.}
\newblock \bibinfo{title}{Experimental evidence for a photon anticorrelation
  effect on a beam splitter: a new light on single-photon interferences}.
\newblock \emph{\bibinfo{journal}{EPL-Europhys. Lett.}}
  \textbf{\bibinfo{volume}{1}}, \bibinfo{pages}{173} (\bibinfo{year}{1986}).

\bibitem{PhysRevLett.59.2044}
\bibinfo{author}{Hong, C.~K.}, \bibinfo{author}{Ou, Z.~Y.} \&
  \bibinfo{author}{Mandel, L.}
\newblock \bibinfo{title}{Measurement of subpicosecond time intervals between
  two photons by interference}.
\newblock \emph{\bibinfo{journal}{Phys. Rev. Lett.}}
  \textbf{\bibinfo{volume}{59}}, \bibinfo{pages}{2044--2046}
  (\bibinfo{year}{1987}).

\bibitem{PhysRevLett.67.318}
\bibinfo{author}{Zou, X.~Y.}, \bibinfo{author}{Wang, L.~J.} \&
  \bibinfo{author}{Mandel, L.}
\newblock \bibinfo{title}{Induced coherence and indistinguishability in optical
  interference}.
\newblock \emph{\bibinfo{journal}{Phys. Rev. Lett.}}
  \textbf{\bibinfo{volume}{67}}, \bibinfo{pages}{318--321}
  (\bibinfo{year}{1991}).

\bibitem{PhysRevA.44.4614}
\bibinfo{author}{Wang, L.~J.}, \bibinfo{author}{Zou, X.~Y.} \&
  \bibinfo{author}{Mandel, L.}
\newblock \bibinfo{title}{Induced coherence without induced emission}.
\newblock \emph{\bibinfo{journal}{Phys. Rev. A}} \textbf{\bibinfo{volume}{44}},
  \bibinfo{pages}{4614--4622} (\bibinfo{year}{1991}).

\bibitem{PhysRevLett.72.629}
\bibinfo{author}{Herzog, T.~J.}, \bibinfo{author}{Rarity, J.~G.},
  \bibinfo{author}{Weinfurter, H.} \& \bibinfo{author}{Zeilinger, A.}
\newblock \bibinfo{title}{Frustrated two-photon creation via interference}.
\newblock \emph{\bibinfo{journal}{Phys. Rev. Lett.}}
  \textbf{\bibinfo{volume}{72}}, \bibinfo{pages}{629--632}
  (\bibinfo{year}{1994}).

\bibitem{PhysRevLett.81.5039}
\bibinfo{author}{Weihs, G.}, \bibinfo{author}{Jennewein, T.},
  \bibinfo{author}{Simon, C.}, \bibinfo{author}{Weinfurter, H.} \&
  \bibinfo{author}{Zeilinger, A.}
\newblock \bibinfo{title}{Violation of bell's inequality under strict einstein
  locality conditions}.
\newblock \emph{\bibinfo{journal}{Phys. Rev. Lett.}}
  \textbf{\bibinfo{volume}{81}}, \bibinfo{pages}{5039--5043}
  (\bibinfo{year}{1998}).

\bibitem{RevModPhys.86.419}
\bibinfo{author}{Brunner, N.}, \bibinfo{author}{Cavalcanti, D.},
  \bibinfo{author}{Pironio, S.}, \bibinfo{author}{Scarani, V.} \&
  \bibinfo{author}{Wehner, S.}
\newblock \bibinfo{title}{Bell nonlocality}.
\newblock \emph{\bibinfo{journal}{Rev. Mod. Phys.}}
  \textbf{\bibinfo{volume}{86}}, \bibinfo{pages}{419--478}
  (\bibinfo{year}{2014}).

\bibitem{PhysRevA.59.1070}
\bibinfo{author}{Bennett, C.~H.} \emph{et~al.}
\newblock \bibinfo{title}{Quantum nonlocality without entanglement}.
\newblock \emph{\bibinfo{journal}{Phys. Rev. A}} \textbf{\bibinfo{volume}{59}},
  \bibinfo{pages}{1070--1091} (\bibinfo{year}{1999}).

\bibitem{gu2019quantum}
\bibinfo{author}{Gu, X.}, \bibinfo{author}{Erhard, M.},
  \bibinfo{author}{Zeilinger, A.} \& \bibinfo{author}{Krenn, M.}
\newblock \bibinfo{title}{Quantum experiments and graphs ii: Quantum
  interference, computation, and state generation}.
\newblock \emph{\bibinfo{journal}{Proc. Nat. Acad. Sci.}}
  \textbf{\bibinfo{volume}{116}}, \bibinfo{pages}{4147--4155}
  (\bibinfo{year}{2019}).

\bibitem{lemos2014quantum}
\bibinfo{author}{Lemos, G.~B.} \emph{et~al.}
\newblock \bibinfo{title}{Quantum imaging with undetected photons}.
\newblock \emph{\bibinfo{journal}{Nature}} \textbf{\bibinfo{volume}{512}},
  \bibinfo{pages}{409--412} (\bibinfo{year}{2014}).

\bibitem{kalashnikov2016infrared}
\bibinfo{author}{Kalashnikov, D.~A.}, \bibinfo{author}{Paterova, A.~V.},
  \bibinfo{author}{Kulik, S.~P.} \& \bibinfo{author}{Krivitsky, L.~A.}
\newblock \bibinfo{title}{Infrared spectroscopy with visible light}.
\newblock \emph{\bibinfo{journal}{Nat. Photonics}}
  \textbf{\bibinfo{volume}{10}}, \bibinfo{pages}{98--101}
  (\bibinfo{year}{2016}).

\bibitem{paterova2017nonlinear}
\bibinfo{author}{Paterova, A.}, \bibinfo{author}{Lung, S.},
  \bibinfo{author}{Kalashnikov, D.~A.} \& \bibinfo{author}{Krivitsky, L.~A.}
\newblock \bibinfo{title}{Nonlinear infrared spectroscopy free from spectral
  selection}.
\newblock \emph{\bibinfo{journal}{Sci. Rep.}} \textbf{\bibinfo{volume}{7}},
  \bibinfo{pages}{1--8} (\bibinfo{year}{2017}).

\bibitem{paterova2018measurement}
\bibinfo{author}{Paterova, A.}, \bibinfo{author}{Yang, H.},
  \bibinfo{author}{An, C.}, \bibinfo{author}{Kalashnikov, D.} \&
  \bibinfo{author}{Krivitsky, L.}
\newblock \bibinfo{title}{Measurement of infrared optical constants with
  visible photons}.
\newblock \emph{\bibinfo{journal}{New J. Phys.}} \textbf{\bibinfo{volume}{20}},
  \bibinfo{pages}{043015} (\bibinfo{year}{2018}).

\bibitem{paterova2018tunable}
\bibinfo{author}{Paterova, A.~V.}, \bibinfo{author}{Yang, H.},
  \bibinfo{author}{An, C.}, \bibinfo{author}{Kalashnikov, D.~A.} \&
  \bibinfo{author}{Krivitsky, L.~A.}
\newblock \bibinfo{title}{Tunable optical coherence tomography in the infrared
  range using visible photons}.
\newblock \emph{\bibinfo{journal}{Quantum Science and Technology}}
  \textbf{\bibinfo{volume}{3}}, \bibinfo{pages}{025008} (\bibinfo{year}{2018}).

\bibitem{PhysRevLett.118.080401}
\bibinfo{author}{Krenn, M.}, \bibinfo{author}{Hochrainer, A.},
  \bibinfo{author}{Lahiri, M.} \& \bibinfo{author}{Zeilinger, A.}
\newblock \bibinfo{title}{Entanglement by path identity}.
\newblock \emph{\bibinfo{journal}{Phys. Rev. Lett.}}
  \textbf{\bibinfo{volume}{118}}, \bibinfo{pages}{080401}
  (\bibinfo{year}{2017}).

\bibitem{su2019versatile}
\bibinfo{author}{Su, J.} \emph{et~al.}
\newblock \bibinfo{title}{Versatile and precise quantum state engineering by
  using nonlinear interferometers}.
\newblock \emph{\bibinfo{journal}{Opt. Express}} \textbf{\bibinfo{volume}{27}},
  \bibinfo{pages}{20479--20492} (\bibinfo{year}{2019}).

\bibitem{paterova2020hyperspectral}
\bibinfo{author}{Paterova, A.~V.}, \bibinfo{author}{Maniam, S.~M.},
  \bibinfo{author}{Yang, H.}, \bibinfo{author}{Grenci, G.} \&
  \bibinfo{author}{Krivitsky, L.~A.}
\newblock \bibinfo{title}{Hyperspectral infrared microscopy with visible
  light}.
\newblock \emph{\bibinfo{journal}{Sci. Adv.}} \textbf{\bibinfo{volume}{6}},
  \bibinfo{pages}{eabd0460} (\bibinfo{year}{2020}).

\bibitem{kviatkovsky2020microscopy}
\bibinfo{author}{Kviatkovsky, I.}, \bibinfo{author}{Chrzanowski, H.~M.},
  \bibinfo{author}{Avery, E.~G.}, \bibinfo{author}{Bartolomaeus, H.} \&
  \bibinfo{author}{Ramelow, S.}
\newblock \bibinfo{title}{Microscopy with undetected photons in the
  mid-infrared}.
\newblock \emph{\bibinfo{journal}{Sci. Adv.}} \textbf{\bibinfo{volume}{6}},
  \bibinfo{pages}{eabd0264} (\bibinfo{year}{2020}).

\bibitem{topfer2022quantum}
\bibinfo{author}{T{\"o}pfer, S.} \emph{et~al.}
\newblock \bibinfo{title}{Quantum holography with undetected light}.
\newblock \emph{\bibinfo{journal}{Science advances}}
  \textbf{\bibinfo{volume}{8}}, \bibinfo{pages}{eabl4301}
  (\bibinfo{year}{2022}).

\bibitem{hochrainer2021quantum}
\bibinfo{author}{Hochrainer, A.}, \bibinfo{author}{Lahiri, M.},
  \bibinfo{author}{Erhard, M.}, \bibinfo{author}{Krenn, M.} \&
  \bibinfo{author}{Zeilinger, A.}
\newblock \bibinfo{title}{Quantum indistinguishability by path identity and with undetected photons}.
\newblock \emph{\bibinfo{journal}{Rev. Mod. Phys.}} \textbf{\bibinfo{volume}{94}},
  \bibinfo{pages}{025007} (\bibinfo{year}{2022}).


\bibitem{PhysRevA.33.4033}
\bibinfo{author}{Yurke, B.}, \bibinfo{author}{McCall, S.~L.} \&
  \bibinfo{author}{Klauder, J.~R.}
\newblock \bibinfo{title}{Su(2) and su(1,1) interferometers}.
\newblock \emph{\bibinfo{journal}{Phys. Rev. A}} \textbf{\bibinfo{volume}{33}},
  \bibinfo{pages}{4033--4054} (\bibinfo{year}{1986}).

\bibitem{chekhova2016nonlinear}
\bibinfo{author}{Chekhova, M.} \& \bibinfo{author}{Ou, Z.}
\newblock \bibinfo{title}{Nonlinear interferometers in quantum optics}.
\newblock \emph{\bibinfo{journal}{Advances in Optics and Photonics}}
  \textbf{\bibinfo{volume}{8}}, \bibinfo{pages}{104--155}
  (\bibinfo{year}{2016}).

\bibitem{ou2020quantum}
\bibinfo{author}{Ou, Z.} \& \bibinfo{author}{Li, X.}
\newblock \bibinfo{title}{Quantum su (1, 1) interferometers: Basic principles
  and applications}.
\newblock \emph{\bibinfo{journal}{APL Photonics}} \textbf{\bibinfo{volume}{5}},
  \bibinfo{pages}{080902} (\bibinfo{year}{2020}).

\bibitem{herzog1994herzog}
\bibinfo{author}{Herzog, T.}, \bibinfo{author}{Rarity, J.},
  \bibinfo{author}{Weinfurter, H.} \& \bibinfo{author}{Zeilinger, A.}
\newblock \bibinfo{title}{Herzog et al. reply}.
\newblock \emph{\bibinfo{journal}{Phys. Rev. Lett.}}
  \textbf{\bibinfo{volume}{73}}, \bibinfo{pages}{3041} (\bibinfo{year}{1994}).

\bibitem{lu2020three}
\bibinfo{author}{Lu, L.} \emph{et~al.}
\newblock \bibinfo{title}{Three-dimensional entanglement on a silicon chip}.
\newblock \emph{\bibinfo{journal}{Npj Quantum Inf.}}
  \textbf{\bibinfo{volume}{6}}, \bibinfo{pages}{1--9} (\bibinfo{year}{2020}).

\bibitem{PhysRevX.11.031044}
\bibinfo{author}{Krenn, M.}, \bibinfo{author}{Kottmann, J.~S.},
  \bibinfo{author}{Tischler, N.} \& \bibinfo{author}{Aspuru-Guzik, A.}
\newblock \bibinfo{title}{Conceptual understanding through efficient automated
  design of quantum optical experiments}.
\newblock \emph{\bibinfo{journal}{Phys. Rev. X}} \textbf{\bibinfo{volume}{11}},
  \bibinfo{pages}{031044} (\bibinfo{year}{2021}).

\bibitem{PhysRevLett.119.240403}
\bibinfo{author}{Krenn, M.}, \bibinfo{author}{Gu, X.} \&
  \bibinfo{author}{Zeilinger, A.}
\newblock \bibinfo{title}{Quantum experiments and graphs: Multiparty states as
  coherent superpositions of perfect matchings}.
\newblock \emph{\bibinfo{journal}{Phys. Rev. Lett.}}
  \textbf{\bibinfo{volume}{119}}, \bibinfo{pages}{240403}
  (\bibinfo{year}{2017}).

\bibitem{greenberger1989going}
\bibinfo{author}{Greenberger, D.~M.}, \bibinfo{author}{Horne, M.~A.} \&
  \bibinfo{author}{Zeilinger, A.}
\newblock \bibinfo{title}{Going beyond bell’s theorem}.
\newblock In \emph{\bibinfo{booktitle}{Bell’s theorem, quantum theory and
  conceptions of the universe}}, \bibinfo{pages}{69--72}
  (\bibinfo{publisher}{Springer}, \bibinfo{year}{1989}).

\bibitem{greenberger1990bell}
\bibinfo{author}{Greenberger, D.~M.}, \bibinfo{author}{Horne, M.~A.},
  \bibinfo{author}{Shimony, A.} \& \bibinfo{author}{Zeilinger, A.}
\newblock \bibinfo{title}{Bell’s theorem without inequalities}.
\newblock \emph{\bibinfo{journal}{Am. J. Phys.}} \textbf{\bibinfo{volume}{58}},
  \bibinfo{pages}{1131--1143} (\bibinfo{year}{1990}).

\bibitem{Niu:08}
\bibinfo{author}{Niu, X.-L.}, \bibinfo{author}{Huang, Y.-F.},
  \bibinfo{author}{Xiang, G.-Y.}, \bibinfo{author}{Guo, G.-C.} \&
  \bibinfo{author}{Ou, Z.~Y.}
\newblock \bibinfo{title}{Beamlike high-brightness source of
  polarization-entangled photon pairs}.
\newblock \emph{\bibinfo{journal}{Opt. Lett.}} \textbf{\bibinfo{volume}{33}},
  \bibinfo{pages}{968--970} (\bibinfo{year}{2008}).

\bibitem{PhysRevLett.117.210502}
\bibinfo{author}{Wang, X.-L.} \emph{et~al.}
\newblock \bibinfo{title}{Experimental ten-photon entanglement}.
\newblock \emph{\bibinfo{journal}{Phys. Rev. Lett.}}
  \textbf{\bibinfo{volume}{117}}, \bibinfo{pages}{210502}
  (\bibinfo{year}{2016}).

\bibitem{PhysRevLett.77.1917}
\bibinfo{author}{Pittman, T.~B.} \emph{et~al.}
\newblock \bibinfo{title}{Can two-photon interference be considered the
  interference of two photons?}
\newblock \emph{\bibinfo{journal}{Phys. Rev. Lett.}}
  \textbf{\bibinfo{volume}{77}}, \bibinfo{pages}{1917--1920}
  (\bibinfo{year}{1996}).

\bibitem{wiseman2000induced}
\bibinfo{author}{Wiseman, H.} \& \bibinfo{author}{M{\o}lmer, K.}
\newblock \bibinfo{title}{Induced coherence with and without induced emission}.
\newblock \emph{\bibinfo{journal}{Phys. Lett. A.}}
  \textbf{\bibinfo{volume}{270}}, \bibinfo{pages}{245--248}
  (\bibinfo{year}{2000}).

\bibitem{PhysRevA.100.053839}
\bibinfo{author}{Lahiri, M.}, \bibinfo{author}{Hochrainer, A.},
  \bibinfo{author}{Lapkiewicz, R.}, \bibinfo{author}{Lemos, G.~B.} \&
  \bibinfo{author}{Zeilinger, A.}
\newblock \bibinfo{title}{Nonclassicality of induced coherence without induced
  emission}.
\newblock \emph{\bibinfo{journal}{Phys. Rev. A}}
  \textbf{\bibinfo{volume}{100}}, \bibinfo{pages}{053839}
  (\bibinfo{year}{2019}).

\bibitem{rudolph2017optimistic}
\bibinfo{author}{Rudolph, T.}
\newblock \bibinfo{title}{Why i am optimistic about the silicon-photonic route
  to quantum computing}.
\newblock \emph{\bibinfo{journal}{APL Photonics}} \textbf{\bibinfo{volume}{2}},
  \bibinfo{pages}{030901} (\bibinfo{year}{2017}).

\bibitem{Feng:23}
\bibinfo{author}{Feng, L.-T.} \emph{et~al.}
\newblock \bibinfo{title}{On-chip quantum interference between the origins of a
  multi-photon state}.
\newblock \emph{\bibinfo{journal}{Optica}} \textbf{\bibinfo{volume}{10}},
  \bibinfo{pages}{105--109} (\bibinfo{year}{2023}).

\end{thebibliography}

\section*{Reference}
\begin{thebibliography}{1}
\expandafter\ifx\csname url\endcsname\relax
  \def\url#1{\texttt{#1}}\fi
\expandafter\ifx\csname urlprefix\endcsname\relax\def\urlprefix{URL }\fi
\providecommand{\bibinfo}[2]{#2}
\providecommand{\eprint}[2][]{\url{#2}}

\bibitem{PhysRevLett.77.1917}
\bibinfo{author}{Pittman, T.~B.} \emph{et~al.}
\newblock \bibinfo{title}{Can two-photon interference be considered the
  interference of two photons?}
\newblock \emph{\bibinfo{journal}{Phys. Rev. Lett.}}
  \textbf{\bibinfo{volume}{77}}, \bibinfo{pages}{1917--1920}
  (\bibinfo{year}{1996}).

\bibitem{gu2019quantum}
\bibinfo{author}{Gu, X.}, \bibinfo{author}{Erhard, M.},
  \bibinfo{author}{Zeilinger, A.} \& \bibinfo{author}{Krenn, M.}
\newblock \bibinfo{title}{Quantum experiments and graphs ii: Quantum
  interference, computation, and state generation}.
\newblock \emph{\bibinfo{journal}{Proc. Nat. Acad. Sci.}}
  \textbf{\bibinfo{volume}{116}}, \bibinfo{pages}{4147--4155}
  (\bibinfo{year}{2019}).

\bibitem{Feng:23}
\bibinfo{author}{Feng, L.-T.} \emph{et~al.}
\newblock \bibinfo{title}{On-chip quantum interference between the origins of a
  multi-photon state}.
\newblock \emph{\bibinfo{journal}{Optica}} \textbf{\bibinfo{volume}{10}},
  \bibinfo{pages}{105--109} (\bibinfo{year}{2023}).

\bibitem{scheidl2010violation}
\bibinfo{author}{Scheidl, T.} \emph{et~al.}
\newblock \bibinfo{title}{Violation of local realism with freedom of choice}.
\newblock \emph{\bibinfo{journal}{Proc. Nat. Acad. Sci.}}
  \textbf{\bibinfo{volume}{107}}, \bibinfo{pages}{19708--19713}
  (\bibinfo{year}{2010}).

\bibitem{larsson2014loopholes}
\bibinfo{author}{Larsson, J.-{\AA}.}
\newblock \bibinfo{title}{Loopholes in bell inequality tests of local realism}.
\newblock \emph{\bibinfo{journal}{J. Phys. A Math. Theor.}}
  \textbf{\bibinfo{volume}{47}}, \bibinfo{pages}{424003}
  (\bibinfo{year}{2014}).

\end{thebibliography}

\section*{Acknowledgements}
This research was supported by the National Key Research and Development Program of China (Grants No. 2022YFE0137000, No. 2019YFA0308704, and No. 2017YFA0303704), the National Natural Science Foundation of China (Grants No. 11690032 and No. 11321063), the NSFC-BRICS (Grant No. 61961146001), the Leading-Edge Technology Program of Jiangsu Natural Science Foundation (Grant No. BK20192001), the Fundamental Research Funds for the Central Universities, the Innovation Program for Quantum Science and Technology (Grant No. 2021ZD0301500), and the Jiangsu Funding Program for Excellent Postdoctoral Talent (No. 20220ZB60).

\section*{Author Contributions} 
K.Q., K.W., L.C. and Z.H. performed the experiment. K.Q., X.-s.M. analysed the data. K.W., M.K. and X.-s.M. designed the experiment. K.W., M.K. and X.-s.M. wrote the manuscript with input from all authors. S.Z. and X.-s.M. supervised and directed the project. All authors commented on the manuscript.

\section*{Supplementary Note}
\subsection*{Supplementary Note 1 -- Four-photon frustrated interference}
In this section, we give the equations used in the main text.
We consider the Hamiltonian of spontaneous parametric down-conversion(SPDC)
\begin{equation}
\hat{H}=i\eta(\hat{p}\hat{a}^{\dagger}\hat{b}^{\dagger}-p^{\dagger}\hat{a}\hat{b})
\end{equation}
as $\eta\ll1$, we use the approximate transformation of SPDC:
\begin{align}
\nonumber U_{\rm SPDC}=e^{-i\hat{H}t}&=\hat{I}+g(\hat{p}\hat{a}^{\dagger}\hat{b}^{\dagger}-p^{\dagger}\hat{a}\hat{b})+O(g^2)\\
&\approx \hat{I}+g(\hat{p}\hat{a}^{\dagger}\hat{b}^{\dagger}-p^{\dagger}\hat{a}\hat{b}),
\end{align}
where $g=\eta t$. Here we omit the two-pair generation process in the same crystal, as the coincidence would be covered by the high intensity of single pair (shown below). With Supplementary Eq.(2), The state after crystals I and II are
\begin{align}
\nonumber \ket{\psi_1}&=U_{\rm II}U_{\rm I}\ket{\alpha,\alpha,vac}\\
&=[\hat{I}+g\alpha\hat{a_1}^{\dagger}\hat{a_4}^{\dagger}]_{\rm II}[\hat{I}+g\alpha\hat{a_2}^{\dagger}\hat{a_3}^{\dagger}]_{\rm I}\ket{\alpha,\alpha,vac}
\end{align}
where $\ket{\alpha,\alpha,vac}$ represents the initial two coherent pump states and the vacuum in the down-converted modes. The pump, signals, and idlers will experience different phase shifts: 
\begin{align}
\nonumber \ket{\psi_2}&=U^{\phi_{s1}}U^{\phi_{i1}}U^{\phi_{s2}}U^{\phi_{i2}}U^{\phi_{p1}}U^{\phi_{p2}}\ket{\psi_1}\\
&=[\hat{I}+g\alpha e^{i(\phi_{i2}+\phi_{s2})}\hat{a_1}^{\dagger}\hat{a_4}^{\dagger}]_{\rm II}[\hat{I}+g\alpha e^{i(\phi_{i1}+\phi_{s1})}\hat{a_2}^{\dagger}\hat{a_3}^{\dagger}]_{\rm I}\ket{\alpha e^{i\phi_{p1}},\alpha e^{i\phi_{p2}},vac}
\end{align}
where $U^\phi=e^{i\phi a^{\dagger}a}$. After crystals III and IV, the final state is:
\begin{align}
\nonumber \ket{\psi_f}=&U_{\rm III}U_{\rm IV}\ket{\psi_{2}}\\
\nonumber =&[\hat{I}+g(\alpha e^{i\phi_{p1}}\hat{a_1}^{\dagger}\hat{a_2}^{\dagger}-\hat{p_1}^{\dagger}\hat{a_1}\hat{a_2})]_{\rm III}[\hat{I}+g(\alpha e^{i\phi_{p2}}\hat{a_3}^{\dagger}\hat{a_4}^{\dagger}-\hat{p_2}^{\dagger}\hat{a_3}\hat{a_4})]_{\rm IV}\\
&[\hat{I}+g\alpha e^{i(\phi_{i2}+\phi_{s2})}\hat{a_1}^{\dagger}\hat{a_4}^{\dagger}]_{\rm II}[\hat{I}+g\alpha e^{i(\phi_{i1}+\phi_{s1})}\hat{a_2}^{\dagger}\hat{a_3}^{\dagger}]_{\rm I}\ket{\alpha e^{i\phi_{p1}},\alpha e^{i\phi_{p2}},vac}
\end{align}
Again, we omit the terms with $O(g^2)$ and get
\begin{align}
\nonumber \ket{\psi_f}=& \hat{I}+g\alpha e^{i\phi_{p1}}\hat{a_1}^{\dagger}\hat{a_2}^{\dagger}+g\alpha e^{i\phi_{p2}}\hat{a_3}^{\dagger}\hat{a_4}^{\dagger}+g\alpha e^{i(\phi_{i2}+\phi_{s2})}\hat{a_1}^{\dagger}\hat{a_4}^{\dagger}+g\alpha e^{i(\phi_{i1}+\phi_{s1})}\hat{a_2}^{\dagger}\hat{a_3}^{\dagger}\\
\nonumber & +g^2\alpha^2[e^{i(\phi_{i1}+\phi_{s1}+\phi_{i2}+\phi_{s2})}\hat{a_1}^{\dagger}\hat{a_2}^{\dagger}\hat{a_3}^{\dagger}\hat{a_4}^{\dagger}+e^{i(\phi_{p1}+\phi_{p2})}\hat{a_1}^{\dagger}\hat{a_2}^{\dagger}\hat{a_3}^{\dagger}\hat{a_4}^{\dagger}\\
\nonumber & +e^{i(\phi_{p1}+\phi_{i2}+\phi_{s2})}(\hat{a_1}^{\dagger})^2\hat{a_2}^{\dagger}\hat{a_4}^{\dagger}
+e^{i(\phi_{p1}+\phi_{i1}+\phi_{s1})}\hat{a_1}^{\dagger}(\hat{a_2}^{\dagger})^2\hat{a_3}^{\dagger}\\
&+e^{i(\phi_{p2}+\phi_{i2}+\phi_{s2})}\hat{a_1}^{\dagger}\hat{a_3}^{\dagger}(\hat{a_4}^{\dagger})^2
+e^{i(\phi_{p2}+\phi_{i1}+\phi_{s1})}\hat{a_2}^{\dagger}(\hat{a_3}^{\dagger})^2\hat{a_4}^{\dagger}]\ket{\alpha e^{i\phi_{p1}},\alpha e^{i\phi_{p2}},vac}.
\end{align}
Supplementary Eq.(6) could be rewritten as the photon number state of spatial modes 1-4: 
\begin{align}
\nonumber \ket{\psi}=&\ket{\alpha e^{i\phi_{p1}},\alpha e^{i\phi_{p2}}}_{p_1,p_2}\otimes\{\ket{vac}+p[e^{i(\phi_{s1}+\phi_{i1})}\ket{0110}+e^{i(\phi_{s2}+\phi_{i2})}\ket{1001}+e^{i\phi_{p1}}\ket{1100}+e^{i\phi_{p2}}\ket{0011}]\\
\nonumber &+p^2[e^{i(\phi_{i1}+\phi_{s1}+\phi_{i2}+\phi_{s2})}\ket{1111}+e^{i(\phi_{p1}+\phi_{p2})}\ket{1111}\\
&+\sqrt{2}e^{i(\phi_{i1}+\phi_{s1}+\phi_{p1})}\ket{1210}+\sqrt{2}e^{i(\phi_{i1}+\phi_{s1}+\phi_{p2})}\ket{0121}+\sqrt{2}e^{i(\phi_{p1}+\phi_{i2}+\phi_{s2})}\ket{2101}+\sqrt{2}e^{i(\phi_{p2}+\phi_{i2}+\phi_{s2})}\ket{1012}]\}.
\end{align}
When we consider only the down-converted photons and set $\phi_{p1}=\phi_{p2}$, Supplementary Eq.(7) is exactly Eq.(1).

With Eq.(1), the terms that contribute to coincidence of 1, 3, 4 is $p^2e^{i(\phi_{i1}+\phi_{s1}+\phi_{i2}+\phi_{s2})}\ket{1111}$, $p^2e^{i2\phi_p}\ket{1111}$, $\sqrt{2}p^2e^{i(\phi_p+\phi_{i2}+\phi_{s2})}\ket{1012}$. The first two terms interfere and the last term contributes a constant noise. Therefore, the coincidence rate is
\begin{equation}
P_{1,3,4}=p^4|e^{i(\phi_{i1}+\phi_{s1}+\phi_{i2}+\phi_{s2})}+e^{i2\phi_p}|^2+2p^4=p^4[4+2\cos{(\phi_{i1}+\phi_{s1}+\phi_{i2}+\phi_{s2}-2\phi_p)}]
\end{equation}

\subsection*{Supplementary Note 2 -- Two-photon frustrated interference}
To show the four-photon interference, we first build two identical individual interferometers between sources I and III, and II and IV. In Supplementary Fig. 1, we show the interferometer between I and III. The scheme is almost the same as in Fig. 2b, except for the QWP in the Swap module, which is fixed at $0^{\circ}$ instead of $45^{\circ}$ so that i1 and i2 do not exchange their path.

To test the interference of sources I and III, we block the pump P2, fix the position of M2 ($\phi_p$), and perform a coarse scan of the phases of down-converted photons ($\phi_i$, $\phi_{s1}$) until the interference fringe of two sources emerges. As shown in Supplementary Fig. 2(a), the envelope indicates about 0.2-mm coherent length of the down-converted photons. Then we fix M3 ($\phi_{s1}$) at the place where the visibility of interference is maximum and finely tune the phase $\phi_i$ to obtain the interference pattern. The result is shown in Supplementary Fig. 2(b). The visibility of two-photon coincidence is 95.5\%. For sources II and IV, we carry out the same operations. The result is shown in Supplementary Fig. 2(c) with a visibility of 95.0\%. The error bar is smaller than the data point and is not shown here.

The high visibility of two-photon frustrated interference (FI) ensures the path identity, which is essential for observing the interference of four sources. We also observe single-photon FI with high visibilities, showing high-level indistinguishability of the photons on the same path, as shown in Supplementary Fig. 3(a), (b).

\begin{figure}[htbp]
    \includegraphics[width=16cm]{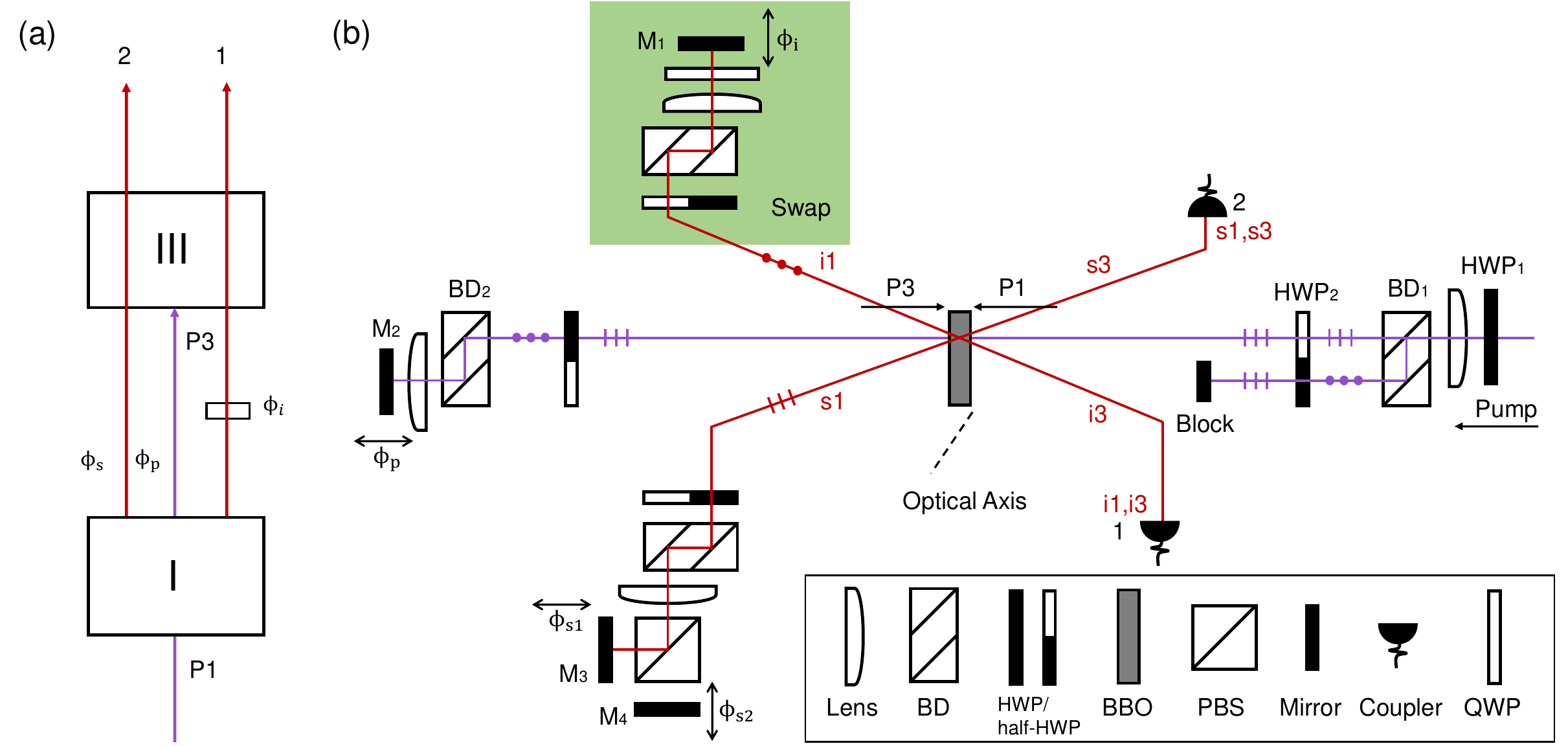}
    \caption{(a) Scheme of frustrated two-photon generation (sources I and III). One pump light pumps two crystals placed in sequence. The signal and idler photons from different sources are aligned to ensure the path identity. The counts of single-photon and two-photon coincidence depend on the phases of pump and down-converted photons: $I\propto1+\cos(\phi_i+\phi_s-\phi_p)$. (b) Experiment setup of the interferometer (sources I and III). We block P2 in Fig. 2b and fix the angle of QWP at $0^{\circ}$ to study the FI of photons from P1 and P3.}
\end{figure}

\begin{figure}[htbp]
    \includegraphics[width=17cm]{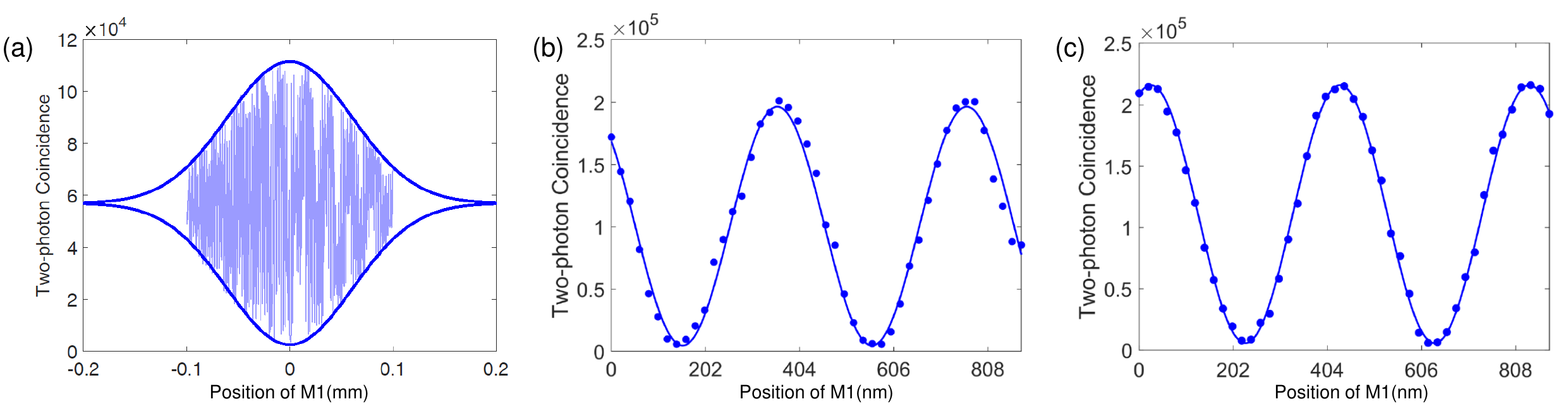}
    \caption{(a) Coarse scan of the phase $\phi_i$ helps to find the locations of interference fringes. (b)/(c) Results of two-fold coincidence counts for the frustrated interference from sources I, III/sources II, IV. The horizontal axis is the position of mirror M1. The interference visibility is (b)/(c) 95.5\%/95.0\%. The integration time of each point is 2~s. }
\end{figure}

\begin{figure}[htbp]
    \includegraphics[width=12cm]{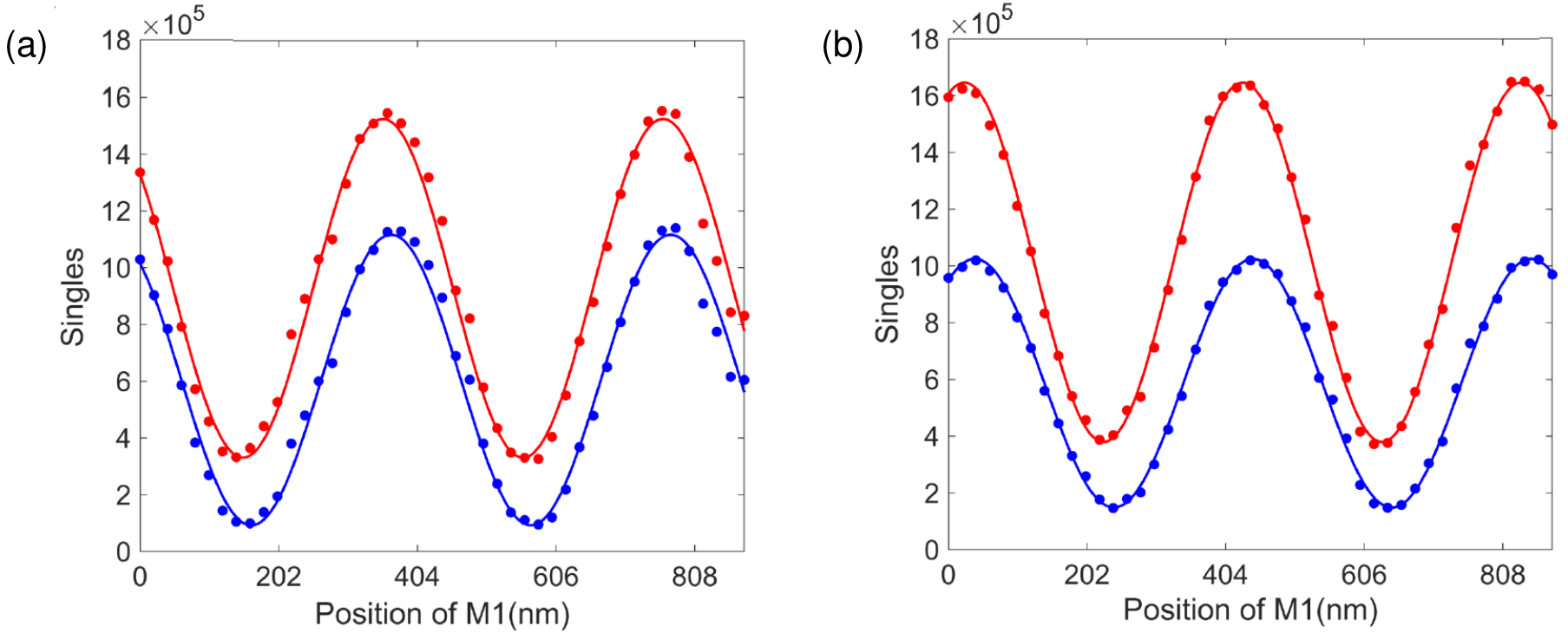}
    \caption{Single-photon counts for the frustrated interference from (a) sources I, III and (b) sources II, IV. The horizontal axis is the position of mirror M1. (a) Blue/red points represent the single counts on detector 1/2 and the visibility of the fitting curve is 84.7\%/64.3\% (b) Blue/red points represent the single counts on detector 3/4 and the visibility of the fitting curve is 74.6\%/62.5\%. The integration time of each point is 2~s.}
\end{figure}

\subsection*{Supplementary Note 3 -- Spatial alignment and interference visibility}
In this section, we discuss the causes of the reduced visibility of FI. We start from the two-photon FI discussed above. The misalignment of photons on the same path (i1/s1 and i3/s3) and the additional loss for photons i1 and s1 on the optical elements give rise to the different coupling efficiency for sources I and III. Therefore, the collected intensity of source I is lower than that of source III, which is the main reason for the limited visibility. This can be seen in Supplementary Table 1, where the two-fold coincidence counts of sources I and II are lower than that of sources III and IV. If we model the different intensity as transmissivity T, the quantum state of sources I and III is
\begin{equation}
\ket{\psi}_{1,3}=p[T_1\ket{11}_{i_1s_1}+R_1\ket{11}_{i'_1s'_1}+\ket{11}_{i_3s_3}],
\end{equation}
where $T_1$ is the transmissivity and $R_1$ is the reflectivity of source I. Photons in the second term are dissipated and not detected. $\ket{11}_{i_1s_1}$ and $\ket{11}_{i_3s_3}$ are indistinguishable and will interfere. The visibility of Supplementary Eq.(9) is ${\rm V}=\frac{2T_1}{1+T_1^2}$, from which we can estimate $T_1$ is about 0.737 with $V=95.5\%$. For sources II and IV, we have a similar analysis and get $T_2=0.724$ with $V=95.0\%$.

\begin{table}
\centering
\begin{tabular}{|p{3cm}|p{3cm}|p{3cm}|p{3cm}|}
    \hline
    \ & \multicolumn{2}{c|}{Coincidence (signal,idler)$\times10^4/s$} & \ \\ \hline
    \ & Before ($N_1$) & After ($N_2$) & $q_i=N_2/N_1$ \\ \hline
    Source I & 1.75 & 1.03 & 0.589 \\ \hline
    Source II & 2.17 & 2.16 & 0.995 \\ \hline
    Source III & 3.27 & 2.43 & 0.743 \\ \hline
    Source IV & 2.83 & 2.34 & 0.827 \\ \hline
\end{tabular}\\
\caption{Two-fold coincidence counts of the four sources before and after the swapping of i1 and i2.}
\end{table}

For the four-photon FI, as the beam spacing and parallelism from the BDs are different, the coupling efficiencies of i1 and i2 reduce significantly after the swapping, which aggravates the intensity imbalance. Therefore, we have to make a compromise between the different sources. The result is shown in Supplementary Table 1. $q_i$ is the ratio of intensity after and before the swapping of source i. Considering that both $T_1$ and $T_2$ above will furthermore reduce the visibility of four-fold coincidence, the quantum state of four-photon FI is
\begin{equation}
\ket{\psi}_{1,2,3,4}=p^2[\sqrt{q_{1}q_{2}}T_{1}T_{2}e^{i(\phi_{i1}+\phi_{s1}+\phi_{i2}+\phi_{s2})}\ket{1111}_{i_{2}s_{1}i_{1}s_{2}}+\sqrt{q_{3}q_{4}}e^{i2\phi_{p}}\ket{1111}_{i_{3}s_{3}i_{4}s_{4}}],
\end{equation}
where the terms that do not contribute to the four-fold coincidence are omitted, and only the two terms that interfere remain. The visibility of Supplementary Eq. (10) is $\rm V=\frac{2\alpha}{1+\alpha^2}=81.9\%$, where $\alpha=\sqrt{\frac{q_1q_2}{q_3q_4}}T_1T_2=0.521$. This estimated visibility from independent two-fold coincidence counts is close to the experiment result of 75.47\% as in the main text. We estimate the reduction of visibility from higher order emission is about 10\% from independent measurement.

For comparison, we reduce $\alpha$ to 0.411, as shown in Supplementary Table 2 and measured the interference pattern for four-photon coincidence counts again. We find the four-photon interference visibility decreases to 67.2\% as shown in Supplementary Fig. 4. The result is consistent with our theoretical prediction $V=70.32\%$ with $\alpha=0.411$.
\begin{table}
\centering
\begin{tabular}{|p{3cm}|p{3cm}|p{3cm}|p{3cm}|}
    \hline
    \ & \multicolumn{2}{c|}{Coincidence (signal,idler)$\times10^4/s$} & \ \\ \hline
    \ & Before ($N_1$) & After ($N_2$) & $q_i=N_2/N_1$ \\ \hline
    Source I & 1.30 & 0.53 & 0.408 \\ \hline
    Source II & 1.21 & 1.41 & 1.17 \\ \hline
    Source III & 2.13 & 2.14 & 1.005 \\ \hline
    Source IV & 2.24 & 1.79 & 0.799 \\ \hline
\end{tabular}\\
\caption{Two-fold coincidence counts of the four sources with lower $\alpha$.}
\end{table}

\begin{figure}[htbp]
    \includegraphics[width=6cm]{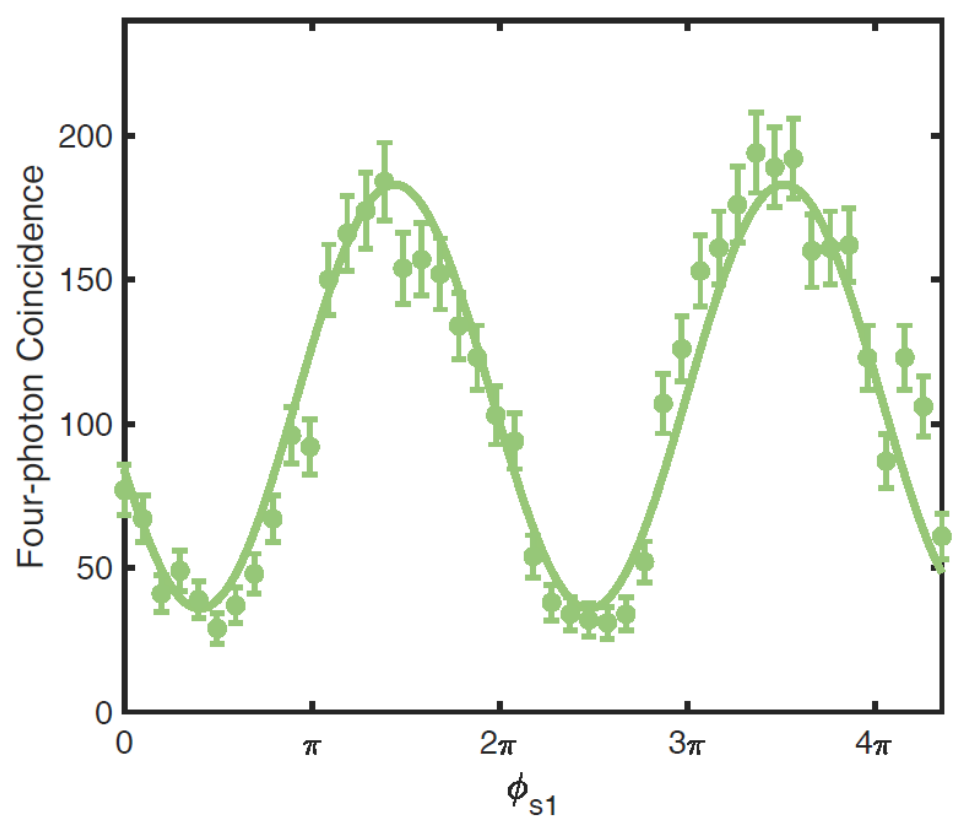}
    \caption{Result of four-fold coincidence counts. The horizontal axis represents the position of M3 ($\phi_{s1}$). The fitting curve has visibility of 67.2\% and a period of 418.9 nm. The error of visibility calculated from Poisson statistics is 4.50\%. }
\end{figure}

\subsection*{Supplementary Note 4 -- Time control of four-photon interference}
The temporal indistinguishability for photons on the same path is essential for the four-photon interference\cite{PhysRevLett.77.1917}. In this section, we analyze the time of the photons generated from different crystals. The lengths of important parts are labeled in Supplementary Fig. 5. 
\begin{figure}[htbp]
    \includegraphics[width=14cm]{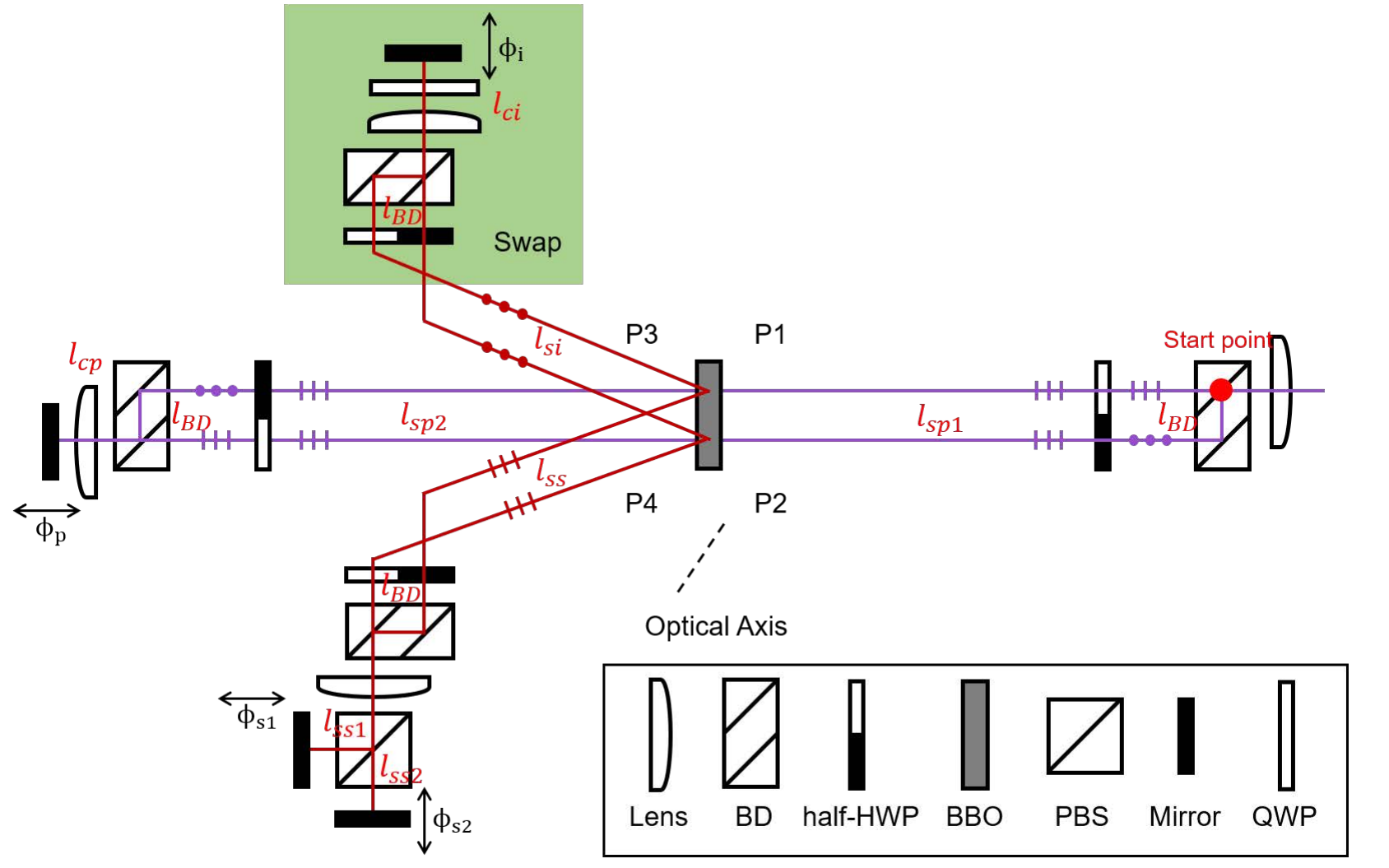}
    \caption{Experiment setup. The lengths of different optical paths are labeled in red. }
\end{figure}

We start our discussion from the two-photon interference scheme, where the QWP is fixed at $0^{\circ}$. The interference of sources I and II occurs when the pump P3 and the photons s1, i1 arrive at the BBO crystal simultaneously. That’s
\begin{equation}
t_{i1}=t_{s1}=t_{P3}
\end{equation}
where
\begin{align}
\nonumber t_{i1}&=\frac{1}{c}(l_{sp1}+l_{si}+l_{BD}+l_{ci}+l_{ci}+l_{BD}+l_{si})\\
&=\frac{1}{c}(l_{sp1}+2l_{si}+2l_{ci}+2l_{BD})\\
\nonumber t_{s1}&=\frac{1}{c}(l_{sp1}+l_{ss}+l_{BD}+l_{ss1}+l_{ss1}+l_{BD}+l_{ss})\\
&=\frac{1}{c}(l_{sp1}+2l_{ss}+2l_{ss1}+2l_{BD})\\
\nonumber t_{P3}&=\frac{1}{c}(l_{sp1}+l_{sp2}+l_{BD}+l_{cp}+l_{cp}+l_{BD}+l_{sp2})\\
&=\frac{1}{c}(l_{sp1}+2l_{sp2}+2l_{cp}+2l_{BD})
\end{align}
represent the time experienced by the photons i1, s1, and pump P3, respectively. The beginning point of time is when the pump incident onto BD1, where the pump splits into two paths, as denoted in Supplementary Fig. 5 with “start point”. As the position of M2 is fixed ($l_{sp2}+l_{cp}$), Supplementary Eq. (11) could be rewritten as
\begin{align}
& t_{i1}=t_{P3} \Rightarrow l_{ci}=l_{sp2}+l_{cp}-l_{si}\\
& t_{s1}=t_{P3} \Rightarrow l_{ss1}=l_{sp2}+l_{cp}-l_{ss}
\end{align}
Thus, we should adjust M1 and M3 to meet the above conditions. For sources II and IV, we have similar conditions:
\begin{align}
&l_{ci}=l_{sp2}+l_{cp}-l_{si}\\
&l_{ss2}=l_{sp2}+l_{cp}-l_{ss}
\end{align}
When we swap i1 and i2, the time the photons experience are as follows:
\begin{align}
&t'_{i1}=\frac{1}{c}(l_{sp1}+2l_{si}+2l_{ci}+l_{BD})\\
&t_{s1}=\frac{1}{c}(l_{sp1}+2l_{ss}+2l_{ss1}+2l_{BD})\\
&t'_{i2}=\frac{1}{c}(l_{sp1}+2l_{si}+2l_{ci}+2l_{BD})\\
&t_{s2}=\frac{1}{c}(l_{sp1}+2l_{ss}+2l_{ss2}+l_{BD})\\
&t_{P3}=\frac{1}{c}(l_{sp1}+2l_{sp2}+2l_{cp}+2l_{BD})\\
&t_{P4}=\frac{1}{c}(l_{sp1}+2l_{sp2}+2l_{cp}+l_{BD})
\end{align}
As long as the conditions of Supplementary Eq. (15) - (18) satisfies, which could be realized by keeping M2, M3, M4 unchanged and scanning the position of M1, the following equations of time indistinguishability still hold:
\begin{align}
&t'_{i2}=t_{P3}\ ({\rm path 1}); \quad t_{s1}=t_{P3}\ ({\rm path 2});\\
&t'_{i1}=t_{P4}\ ({\rm path 3}); \quad t_{s2}=t_{P4}\ ({\rm path 4});
\end{align}
The above analysis shows that, though the photons from the sources I to IV are generated asynchronously\cite{PhysRevLett.77.1917}, we still can’t distinguish the photon on the same path by time, which is essential for the four-photon FI.

\subsection*{Supplementary Note 5 -- Two-fold coincidence counts in the four-photon frustrated interference}
We also analyze the two-fold coincidence counts in the four-photon interference experiment. The result is shown in Supplementary Fig. 6. There are six two-fold coincidences of the four photons on modes 1 to 4. Only the coincidences on detectors 1 and 3 or detectors 2 and 4 will show the interference pattern, as shown in Supplementary Fig. 6(a). The interference is a result of four-photon FI. Affected by the noise from the first order and second order, as shown in Eq. (1), the interference visibility is limited. The coincidence counts of detectors 2, 3 (source I), detectors 1, 4 (source II), detectors 1, 2 (source III) and detectors 3, 4 (source IV) show no interference as shown in Supplementary Fig. 6(b).

\begin{figure}[htbp]
    \includegraphics[width=14cm]{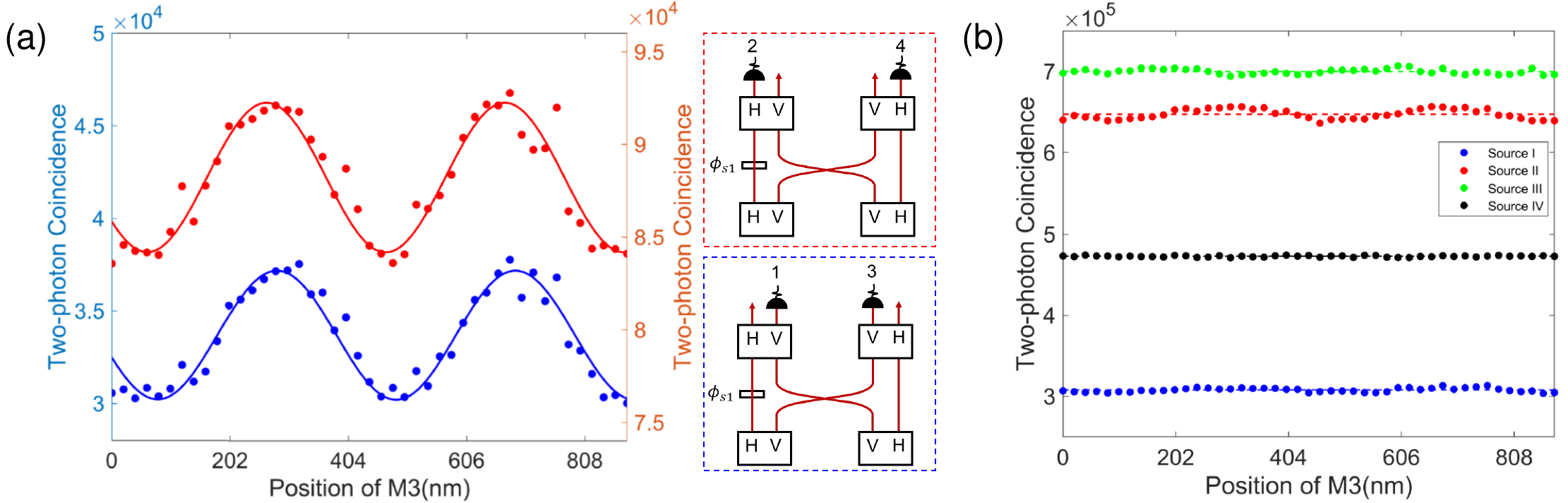}
    \caption{ Results of two-fold coincidence counts in the four-photon frustrated interference. The horizontal axis represents the position of M3. (a) The red line represents the coincidence of detectors 2 and 4. The fitting curve has a visibility of 4.57\% and a period of 407.0 nm. The blue line represents the coincidence of detectors 1 and 3. The fitting curve has a visibility of 10.33\% and a period of 407.2 nm. (b) The coincidence counts of photons from four different sources I (detector 2,3), II (detector 1,4), III (detector 1,2), and IV (detector 3,4) show no interference, as expected.}
\end{figure}

\subsection*{Supplementary Note 6 -- Visibility and transmissivity (V-T)}
In this section, we give the function used for fitting the V-T correlation in the main text. Supplementary Eq. (10) gives the final state collected by detectors 1-4. When we reduce the transmissivity T of photon s2, it should be rewritten as:
\begin{equation}
\ket{\psi}_{1,2,3,4}=\sqrt{q_3q_4}p^2[\alpha Te^{i(\phi_{i1}+\phi_{s1}+\phi_{i2}+\phi_{s2})}\ket{1111}_{i_{3}s_{3}i_{4}s_{4}}+\alpha R e^{i(\phi_{i1}+\phi_{s1}+\phi_{i2}+\phi_{s2})}\ket{1111}_{i_{3}s_{3}i_{4}s_{2}}+e^{i2\phi_{p}}\ket{1111}_{i_{3}s_{3}i_{4}s_{4}}],
\end{equation}
where we have denoted identical photons with the same subscript. R is the reflectivity of photon s2. The second term is dissipated and not detected. Therefore, the second term will not contribute to four-fold coincidence but to coincidences on detectors 1, 2, and 3. The visibility of the above equation for four-fold coincidence is 
\begin{equation}
V=\frac{2\alpha T}{1+(\alpha T)^2}
\end{equation}
We consider $\alpha$ as a variable and are used to quantify the path identity of the photons or the intensity imbalance of the sources, as stated above. Supplementary Fig. 7 shows the four-photon FI with different transmissivities of photon s2, which corresponds to Fig. 3c. As T is reduced, the visibility tends to vanish. The $\alpha$ calculated from coincidence counts (0.521) is higher than that from the fitting curve (0.42) with Supplementary Eq. (28) . The difference may come from the limited long-time stability of our experiment and multiphoton noise from high-order emission.
\begin{figure}[htbp]
    \includegraphics[width=10cm]{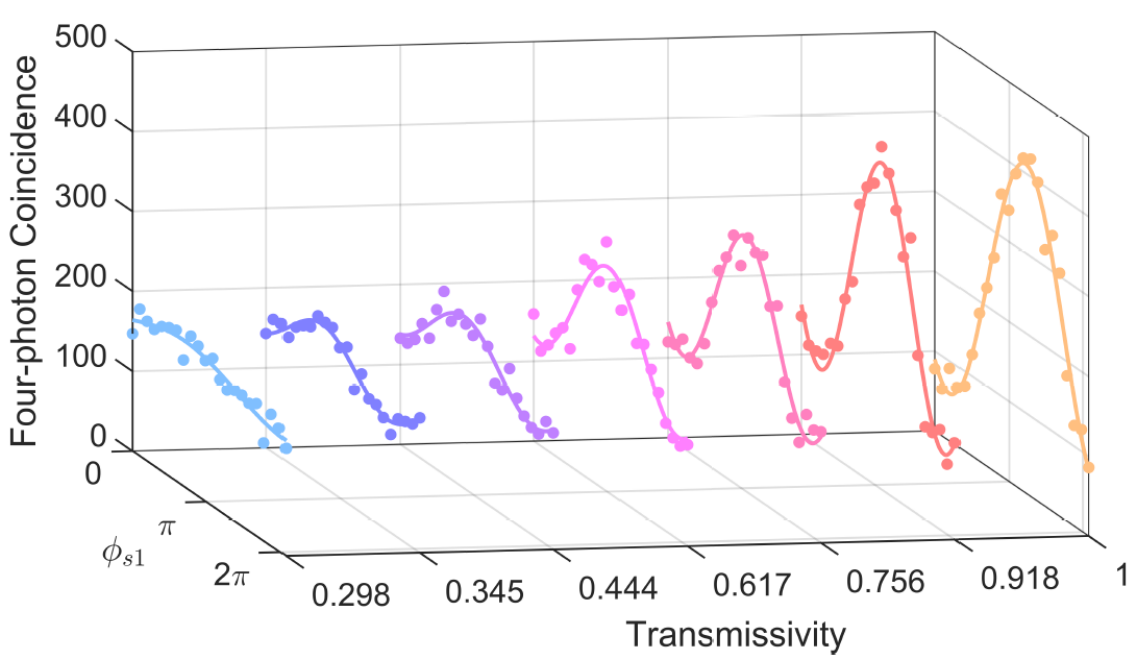}
    \caption{The relationship between the visibility of four-photon coincidence and the transmissivity of photon s2.}
\end{figure}

For the coincidences on detectors 1, 2, and 3, as the second term in Supplementary Eq. (27) and term $\ket{1210}$ in Eq. (1) will contribute a constant noise, the visibility is 
\begin{equation}
V=\frac{2\alpha T}{\alpha^2+3}.
\end{equation}
V is proportion to T. Supplementary Fig. 8 shows the four-photon FI with different transmissivities of photon s2, which corresponds to Fig. 4c. As T is reduced, the visibility again tends to vanish.
\begin{figure}[htbp]
    \includegraphics[width=10cm]{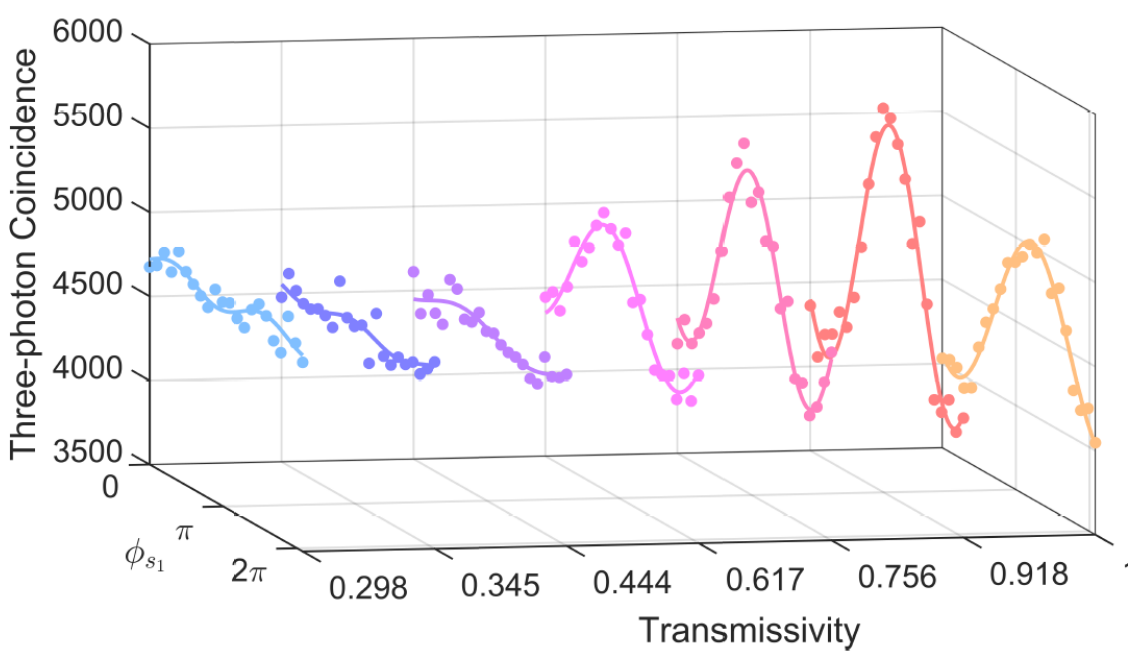}
    \caption{The relationship between the visibility of three-photon coincidence and the transmissivity of photon s2.}
\end{figure}

\subsection*{Supplementary Note 7 -- Space-time diagram of non-local and local quantum interference}
Throughout our manuscript, we use the terms ‘non-local’ by its operational definitions. Following the strict Einstein’s locality conditions, when we say events X and Y are non-local, it means X lies outside of both the future and past lightcones of Y. 

The quantum interference in the frustrated two-photon generation is the interference between photon sources from two crystals. Based on the above definition of non-local, as the twin photons are generated in the same place, the phase setting event \textbf{a} is always in the past light cone of the detection event \textbf{B} (see Supplementary Fig. 9b below). Therefore, we call it local quantum interference. 

The analogy between the two-photon entanglement and four-photon frustrated interference is twofold:

1. From the space-time configuration perspective: While the four-photon frustrated interference expands the source into a larger space by swapping the idlers, the phase-setting event \textbf{a} (phase $\alpha$) can be space-like separated from the detection event \textbf{B}. We call this non-local quantum interference (Supplementary Fig. 9c below). This situation for the four-photon frustrated interference is similar to the two-photon entanglement (Supplementary Fig. 9a below), in which the setting events \textbf{a}/\textbf{b} can be space-like separated from the detection events \textbf{B}/\textbf{A}.

2. From the experimental observation perspective: For an EPR state (Supplementary Fig. 9a below), when Alice and Bob change their measurement settings, their single counts remain constant, while the coincidence between them shows dependence on both settings. In the four-photon frustrated interference (Supplementary Fig. 9c below), when Alice and Bob change their phase settings ($\alpha$ and $\beta$), their local two-photon coincidence counts (photons 1 and 2, photons 3 and 4, respectively) remain constant, while the four-photon coincidence shows the interference depending on both settings. In particular, when Alice and Bob set the phase $\alpha+\beta=\pi$ and keep them unchanged, there will be no four-photon coincidence counts. Therefore, when Alice detects two photons (event \textbf{A}), she can state that there must be no two-photon coincidence counts at Bob’s side (event \textbf{B}); that is, event \textbf{A} prohibits event \textbf{B} with the phase setting $\alpha+\beta=\pi$, no matter how far Lab1 and Lab2 are separated from each other. We call this photon-count conditional probability between Alice and Bob the non-local quantum interference.

\begin{figure}[htbp]
    \includegraphics[width=16cm]{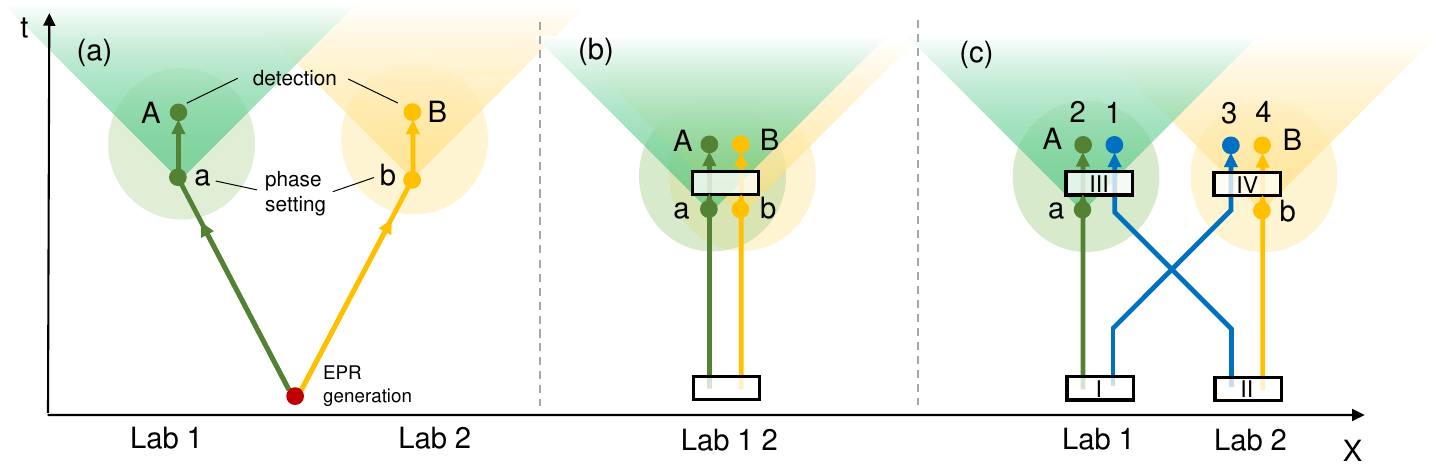}
    \caption{Space-time diagram of the three experiment settings. (a) Entangled state. (b) Two-photon frustrated interference. (c) Four-photon frustrated interference.}
\end{figure}

So far, in our experiment, a single laser is the pump source for all of the photon pair generations -- and this could open the loophole at the time of the creation of the photon pairs in the lower layer, the phase setting is already fixed. However, this is only done for technical reasons. In principle, the phenomenon we have observed could be demonstrated under strict Einstein locality conditions\cite{gu2019quantum,Feng:23}. There are several possible implementations, and we explain two of them now, referring to Fig. 1c:\\
(1) If the pump source is a pulsed laser, then the phases $\alpha$ and $\beta$ can be set randomly after the pump pulse goes through the two crystals in the lower layer.\\
(2) The pump sources could be four independent phase-stable lasers. In that way, again it is possible to separate the four sources and the phases $\alpha$ and $\beta$, such that the phases are set only after the photon pairs are produced in the lower layer.\\

Note this requires that the phase of Alice and Bob should be in the same reference frame of the pump light, which may potentially open certain loopholes, such as the freedom-of-choice loophole, especially when one consider the deterministic model. See details in [PNAS 107, 19708 (2010)]\cite{scheidl2010violation}.  However, to close all loopholes is beyond the scope of the current work and we plan to address this interesting topic in the future. Additionally, investigating the relations of these experimental design to other loopholes will be an interesting theoretical study\cite{larsson2014loopholes}. To close the locality loophole, we can introduce a random phase on the pump, and configure the phase setting event for Alice/Bob outside the light cone of the phase setting event for the pump.

\subsection*{Supplementary Note 8 -- Cancel the noise of three-photon interference}
Here we give an example of source configuration to improve the three-photon interference visibility. In the main text, it is shown that the maximal three-photon interference visibility is 50\% because of the constant noise term $\ket{1,0,1,2}$ produced by a simultaneous event in crystals II and IV.

This, however, is not a fundamental limit, as the additional noise contribution can be canceled by destructive interference (Supplementary Fig. 10). Let’s first understand this via the graph representation of our experiment. As shown in Fig. S10a below, each vertex represents the path mode of photons. Each blue edge represents a photon pair source, (i.e., a non-linear crystal), which is labeled from I to IV as shown in Fig. 2. The noise term reducing the three-photon interference (photons 1, 3, and 4 in our experiment) is $\ket{1,0,1,2}$. So we have to make revisions in the setup in modes 1, 3, and 4. We now add four more crystals: V, VI, VII, and VIII. Crystal V emits non-collinear pairs into modes 1 and 3. Crystal VI produces two photons collinearly in path 4. They together produce an additional three-fold detection in the form $\ket{1,0,1,2}$. If we set the phase of the light that pumps crystal VI to $\pi$ with respect to the light that pumps crystal I, then the noise contribution cancels with the newly created term.

At this stage, two additional terms emerge from a combination of crystal V with crystal II and with crystal IV. However, those contributions can also be canceled by adding new crystals VII and VIII that produce two photons collinearly in paths 1 and 3, respectively. Therefore, all noise contributions cancel and the resulting three-photon interference has 100\% visibility. Detailed terms (on modes 1, 3, 4) generated from newly added crystals are listed below:\\
\textbf{Noise cancel:} Crystal V \& VI: $-\ket{1,0,1,2}$\\
\textbf{Newly created noise terms that have coincidence count:}\\
\quad Crystal V \& II: $\ket{2,0,1,1}$\\
\quad Crystal V \& IV: $\ket{1,0,2,1}$\\
\textbf{Terms from the contribution of Crystals VII and VIII:}\\
\quad Crystal VI \& VII: $-\ket{2,0,1,1}$\\
\quad Crystal II \& VIII: $-\ket{1,0,2,1}$\\
The four above terms cancel.

\begin{figure}[htbp]
    \includegraphics[width=10cm]{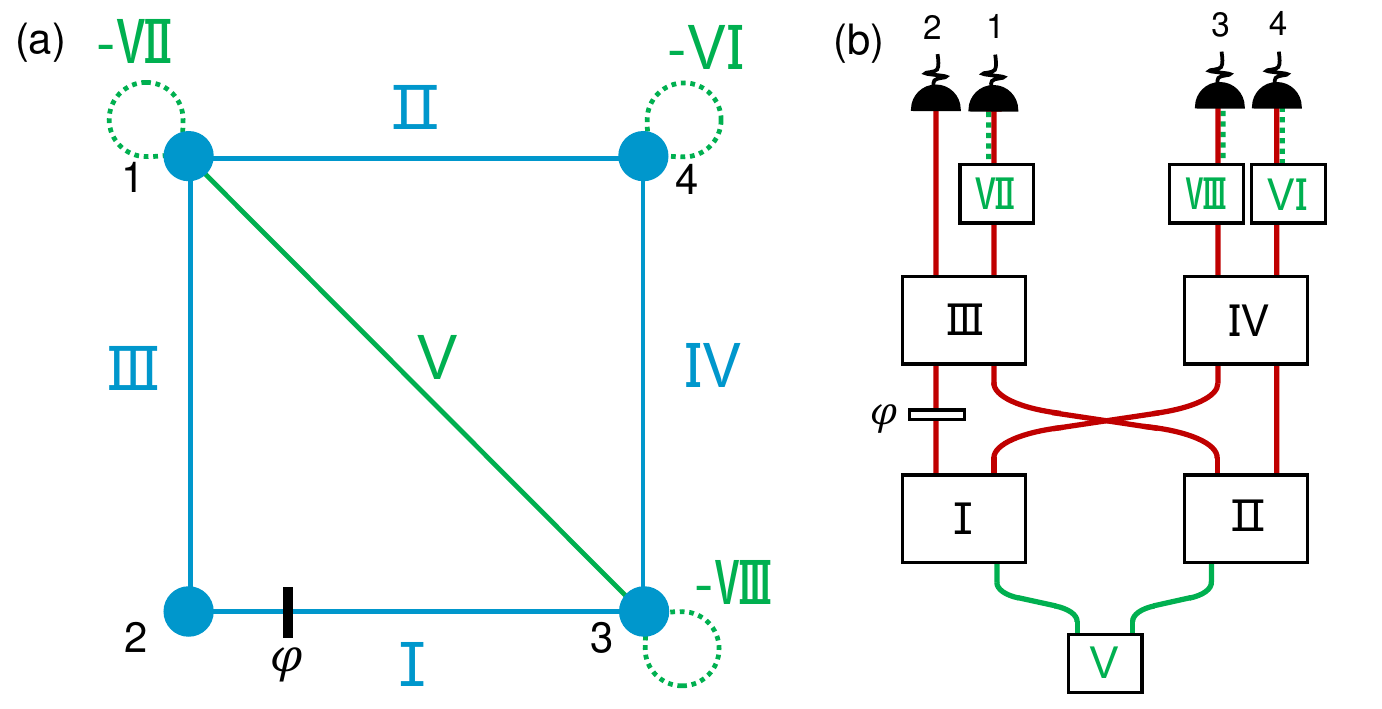}
    \caption{(a) Graph representation of noise cancellation. (b) The noise of three-photon interference can be completely restrained by adding 4 crystals marked in green.}
\end{figure}

\subsection*{Supplementary Note 9 -- The Cause of non-local quantum interference}
As shown in Supplementary Fig. 11, one can generate the GHZ state from a system similar to that of the frustrated four-photon interference, except that we use mode shifters to convert the photons coming from crystals I and II from state $\ket{0000}$ to $\ket{1111}$. In this configuration, the GHZ state $\ket{\Psi_{GHZ}}=\frac{1}{\sqrt{2}}[\ket{1111}+e^{i\varphi}\ket{0000}]$ will not show interference when one traces over the undetected photon (photon 4), as the photons 123 are in a mixed state $\rho_{123}=\frac{1}{2}\ket{000}\bra{000}+\frac{1}{2}\ket{111}\bra{111}$. However, when we remove the mode shifter of the undetected photon (photon 4), the undetected photon remains in $\ket{0}$. The final state becomes a bipartite product state between particles 123 and 4: $\ket{\Psi_{\rm product}}=\frac{1}{\sqrt{2}}[\ket{111}+e^{i\varphi}\ket{000}]\ket{0}$. As the state of photons 123 contains the phase information: $\ket{\Psi}_{123}=\frac{1}{\sqrt{2}}[\ket{111}+e^{i\varphi}\ket{000}]$, the interference shows up with the state $\ket{\Psi_{\rm product}}$ even if we take the trace over the undetected photon (photon 4) and measure photons 1, 2, and 3 in mutually unbiased bases, for example in the $\ket{+++}$ base. Therefore, in this case the non-local quantum interference with undetected photon comes from the path identity rather than entanglement. Only when there are identical modes of the undetected photon, will the phase information ($\varphi$) on the undetected photon be transferred to the other three photons. 

\begin{figure}[htbp]
    \includegraphics[width=10cm]{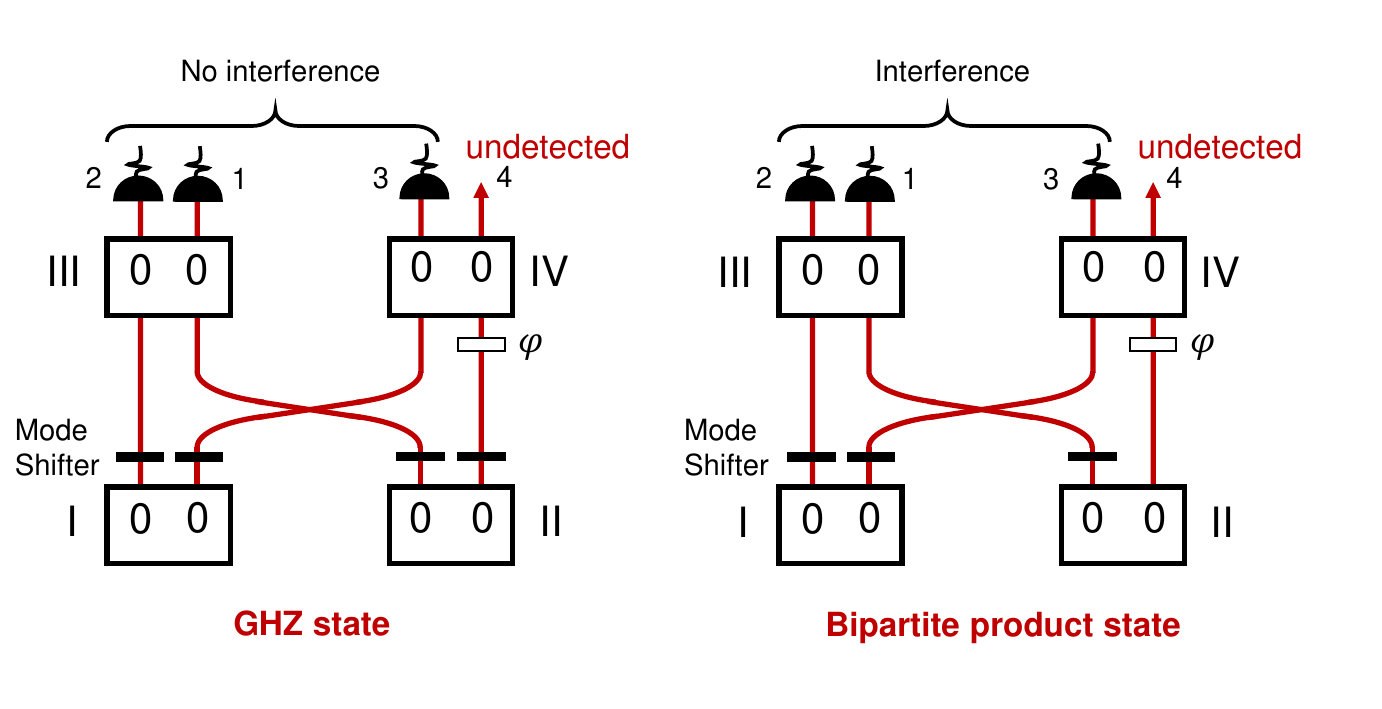}
    \caption{The interference with an undetected photon of the GHZ state and the bipartite product state.}
\end{figure}

\end{document}